\documentclass[aps,prl,twocolumn,superscriptaddress,balancelastpage,showpacs,reprint,english]{revtex4-1}
\usepackage[T1]{fontenc}
\usepackage{amsmath}
\usepackage{amssymb}
\usepackage{commath}
\usepackage{graphicx}
\usepackage{color}
\usepackage{bm}
\usepackage[caption=false]{subfig}
\usepackage{mathrsfs}
\usepackage{hyperref}
\usepackage{babel}
\usepackage[normalem]{ulem}

\DeclareMathAlphabet{\mathdutchcal}{U}{dutchcal}{m}{n}

\usepackage{tikz}
\usetikzlibrary{decorations.pathmorphing, patterns,decorations.text}
\tikzset{snake it/.style={decorate, decoration=snake}}
\usetikzlibrary{shapes.misc}
\tikzset{cross/.style={cross out, draw=black, minimum
size=2*(#1-\pgflinewidth), inner sep=0pt, outer sep=0pt}, 
cross/.default={7pt}}
\tikzset{cross2/.style={cross out, draw=black, minimum
size=2*(#1-\pgflinewidth), inner sep=0pt, outer sep=0pt}, 
cross2/.default={4pt}}

\makeatletter

\newcommand{\qq}{\begin{eqnarray}}
\newcommand{\qqq}{\end{eqnarray}}

\newcommand{\p}{\partial}

\newcommand{\bx}{\mathbf{x}}

\newcommand{\bsfQ}{\bsf Q}

\newcommand{\bsf}[1]{\textsf{\textbf{#1}}}

\makeatother

\begin{document}

\title{Eppur si muove: Shape of topological defects---and consequent motion---in active nematics}
\author{Giacomo Marco La Montagna}
\affiliation{{Laboratoire de Physique Th\'eorique et Mod\'elisation, CNRS UMR 8089,
		CY Cergy Paris Universit\'e, F-95032 Cergy-Pontoise Cedex, France}}
\author{Sumeja Burekovi\'c}
\affiliation{Service de Physique de l'Etat Condens\'e, CEA, CNRS Universit\'e Paris-Saclay, CEA-Saclay, 91191 Gif-sur-Yvette, France}
\affiliation{Sorbonne Universit\'e, CNRS, Laboratoire de Physique Th\'eorique de la Mati\`ere Condens\'ee, 75005 Paris, France}

\author{Ananyo Maitra}
\affiliation{{Laboratoire de Physique Th\'eorique et Mod\'elisation, CNRS UMR 8089,
		CY Cergy Paris Universit\'e, F-95032 Cergy-Pontoise Cedex, France}}
\affiliation{Laboratoire Jean Perrin, Sorbonne Universit\'{e} and CNRS, F-75005, Paris, France}
\author{Cesare Nardini}
\affiliation{Service de Physique de l'Etat Condens\'e, CEA, CNRS Universit\'e Paris-Saclay, CEA-Saclay, 91191 Gif-sur-Yvette, France}
\affiliation{Sorbonne Universit\'e, CNRS, Laboratoire de Physique Th\'eorique de la Mati\`ere Condens\'ee, 75005 Paris, France}

\date{\today}


\begin{abstract}
Topological defects in systems with liquid-crystalline order are crucial in determining their large-scale properties. In active systems, they are known to have properties impossible at equilibrium: for example, $+1/2$ defects in nematically-ordered systems self-propel. While some previous theoretical descriptions relied on assuming that the defect shape remains unperturbed by activity, we show that this assumption can lead to inconsistent predictions. We compute the shape of $-1/2$ defects and show that the one of $+1/2$ is intimately related to their self-propulsion speed.
Our analytical predictions are corroborated via numerical simulations of a generic active nematic theory.
 \end{abstract}

\maketitle

Topological defects are fundamental characteristics of phases displaying liquid-crystalline order \cite{Anderson_cond_matt, Mermin_rev}. In active systems, they have properties that are significantly distinct from their equilibrium counterparts. For instance, in polar active matter, certain defects with charge $+1$ spontaneously rotate \cite{JF_pol_def, Carles_pol1, Carles_pol2, Pol_def_angheluta} and interactions between defects in a variety of symmetry-broken states can be non-reciprocal and non-central \cite{AM_hex, Vincenzo_def, Farzan_nem, Farzan_pol}. In active nematics, defects with charge $+1/2$ are known to self-propel \cite{Vijay_rods, giomi2013, Pismen,  Giomi_def, Doostmohammadi_rev}.

Of all ordered states, topological defects in two-dimensional active nematics have received the most attention \cite{Suraj_rev, Doostmohammadi_rev}. There are at least two reasons for this. First, nematic order is ubiquitous in active materials of biological origin such as those composed of cytoskeletal filaments~\cite{Dogic_nem,Dogic_def} and bacteria~\cite{Sashi_bac,LLC}, in biological tissues \cite{Def_exp3,LLC-cell, Ladoux_act_nem}, and it is also found in artificial active systems, as in monolayers of vibrated granular rods \cite{Vijay_rods}. Second, defects in active nematics dramatically modify the mechanical and statistical properties of the ordered phase.
For example, the quasi-long-range-ordered nematic phase in two dimensions was shown to be re-entrant because, due to the self-propulsion of the $+1/2$ defect, $\pm 1/2$ defects pairs generically unbind at low noise \cite{Suraj_def}. Moreover, in addition to being of fundamental physical importance, defects in active nematics have also been implicated in a variety of biological processes that have functional consequences: it has been argued that they drive shape changes of bacterial colonies \cite{Def_morph1}, have a crucial role in the morphogenesis of hydra \cite{Def_morph2,Def_morph3}, and drive flows \cite{Dogic_def} leading to the spatiotemporally chaotic state known as active turbulence \cite{Doostmohammadi_rev, Luca_PRX}. Further, defects sustain a local density gradient leading to enhanced or reduced density at the core \cite{Shradha, Def_exp1} and were implicated in the control of cell death and extrusion \cite{Def_extru}.

\begin{figure}[h]\center
	\centering
	\includegraphics[width=1.0\linewidth]{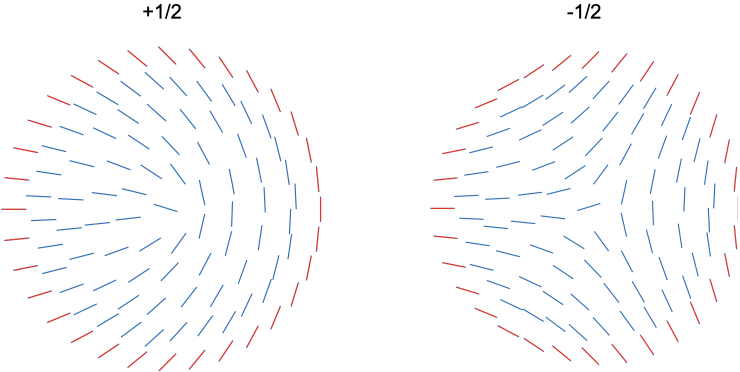}
	\caption{Nematic director ${\bf n}=(\cos\theta,\sin\theta)$ orientation for defects of charges $+1/2$ (left) and $-1/2$ (right) with $\theta=\pm\phi/2+\phi_0$ respectively, where $\phi$ is the polar angle and $\phi_0=\pi/2$, obtained by numerical simulations of eq. \eqref{eq:minimal-nematics} in a circular domain. Red lines denote the director orientation at the boundary that is kept fixed in time; this  is chosen to correspond to the far field of the defect in a single Frank constant liquid crystal, which ensures the presence of a $\pm 1/2$ defect in the bulk.}
	\label{Fig1}
\end{figure}

For these reasons, dynamics of topological defects in active nematics have attracted significant theoretical attention \cite{Suraj_rev, Doostmohammadi_rev, Suraj_def, giomi2013, Pismen, Vijay_rods,Angheluta1, Angheluta2}. The self-propulsion speed of $+1/2$ defects was calculated in a frictionally-screened fluid medium perturbatively at small activity~\cite{Pismen}, and the interactions between $+1/2$ defects and between $\pm 1/2$ defects pairs were characterised, uncovering the presence of torques that are absent in the passive case~\cite{Suraj_def, Angheluta2}. 
Importantly, none of these theories calculated the shape of defects and, at least some works~\cite{Giomi_def, Angheluta2, thijssen2020binding, ronning2022flow, maitra2019spontaneous}, crucially assumed that it is unchanged by activity.
However, it has been known that topological defects in nematic liquid crystals have different shapes depending on whether there is a single or two Frank constants~\cite{Dzyalo,Deem,Anghuluta_Vinals} i.e., whether splay and bend deformations have the same elastic cost or not. It is thus natural to expect that activity induces shape-modifications as well, raising the question of how these change the properties of topological defects. As we show, not accounting for those in active systems has serious consequences, even for the most studied case of self-propulsion of $+1/2$ defects. One of the most popular methodologies for calculation of defect speed, the application of Halperin-Mazenko method \cite{Mazenko2} under the assumption that the defect shape is the same as in single-Frank constant liquid crystals~\cite{Angheluta2, ronning2022flow,matsukiyo2025defect,schimming2025analytical,jacques2023aging}, implies that there are classes of active nematics in which defects \emph{should not} self-propel ~\cite{supp}. In models retaining the fluid velocity $\bf v$, the self-propulsion of the $+1/2$ defect was alternatively obtained as the fluid velocity at the core of the defect, again assuming that the defect has the same shape as in single Frank constant systems \cite{Giomi_def,khoromskaia2017vortex,thijssen2020binding,AM_PNAS,houston2023colloids,maitra2019spontaneous}. Once again, this suggests that $+1/2$ defects do not always self-propel~\cite{AM_PNAS}. Is this true?

In this Letter, we describe---for the first time--- the modification of defect shape due to activity. For simplicity we restrict most of our analysis to two-dimensional dry systems, but we expect analogous results in higher dimensions and in the presence of fluid flows (we partially consider the last case in~\cite{supp}). For $-1/2$ defects, we obtain a closed analytical expression for the defect shape both in the far and near field, while for $+1/2$ defects we link the shape at the core to its self-propulsion speed. We further show that ignoring shape modifications can lead to incorrect results even at arbitrarily small activities, and that defects in active nematics generically self-propel, irrespective of the nature of activity. 
Our analytical results are quantitatively confirmed by direct numerical simulations. 

The generic model of dry active nematics we consider is defined by
\qq\label{eq:minimal-nematics}
\p_t Q_{ij}&=& \mathcal{G}_{ij}\,,\nonumber\\
\mathcal{G}_{ij} &=&
 aQ_{ij} -  b\text{Tr}(\bsfQ^2)Q_{ij} + \kappa\nabla^2 Q_{ij} \nonumber\\
&-&\frac{K_1}{4} \mathcal{L}_{ij}\, \text{Tr}({\bsf Q}^2)
+\frac{K_2}{2} Q_{kl} \mathcal{L}_{ij} Q_{kl}\\
&+& L_1 Q_{kl}\nabla_k\nabla_l Q_{ij}
+L_2 (\nabla_k Q_{kl}) (\nabla_l Q_{ij})\nonumber \nonumber
\qqq
where $\mathcal{L}_{ij} = \nabla_i\nabla_j -(\delta_{ij}/2)\nabla^2$, and 
\begin{equation}
	{\bsf Q}\equiv \frac{S}{2}\begin{pmatrix}\cos2\theta & \sin 2\theta\\\sin 2\theta &-\cos 2\theta\end{pmatrix},
\end{equation}
is the traceless and symmetric nematic tensor specified by its magnitude $S$, and by its angle field  $\theta$ about an arbitrary but fixed axis. Without loss of generality, we choose $a=\kappa=1$, which corresponds to setting non-dimensionalised units with the time-scale $1/a$ and the length-scale $\sqrt{\kappa/a}$. We further choose $b=1$, fixing $S=\sqrt{2}$  in a perfectly ordered state. As shown in~\cite{supp}, the spin-wave theory associated with eq. \eqref{eq:minimal-nematics} corresponds to the one previously studied in the literature up to order $\theta^2$~\cite{mishra2010dynamic,shankar2018low}.

Activity enters in eq. (\ref{eq:minimal-nematics}) via the coefficients $L_1$, $L_2$, $K_1$, $K_2$. Crucially, there are two equilibrium limits of eq. \eqref{eq:minimal-nematics}: if $L_1=L_2=K_1=K_2=0$, it reduces to a passive liquid-crystal with the single Frank constant $\kappa$. When $L_1=L_2=K_1=K_2=K$ instead, eq. (\ref{eq:minimal-nematics}) reduces to an equilibrium liquid-crystal with two Frank constants $\kappa$ and $K$. In this case, we have $\mathcal{G}_{ij}=-\delta \mathcal{F}/\delta Q_{ij}$, with
\qq\label{eq:free-energy-2-frank}
\mathcal{F} 
&=& \frac{1}{2}\int_\bx \Big[-a\text{Tr}(\bsfQ^2) +\frac{b}{2}\text{Tr}(\bsfQ^4)+\kappa (\nabla\bsfQ)^2\nonumber\\
&+&{K} \,Q_{kl} (\nabla_k Q_{ij})\nabla_l Q_{ij})\Big]\,.
\qqq
We will use the single Frank constant and two Frank constant equilibrium limits to critically examine methodologies used in the literature \cite{Pismen, Angheluta2, Angheluta1, Suraj_def} to discuss defect dynamics. We further show that a model of the form of eq.~\eqref{eq:minimal-nematics} can be obtained at large scales starting from a compressible wet active nematics on a substrate~\cite{supp}.

\begin{figure}
	\begin{centering}
				\includegraphics[width=1.\columnwidth]{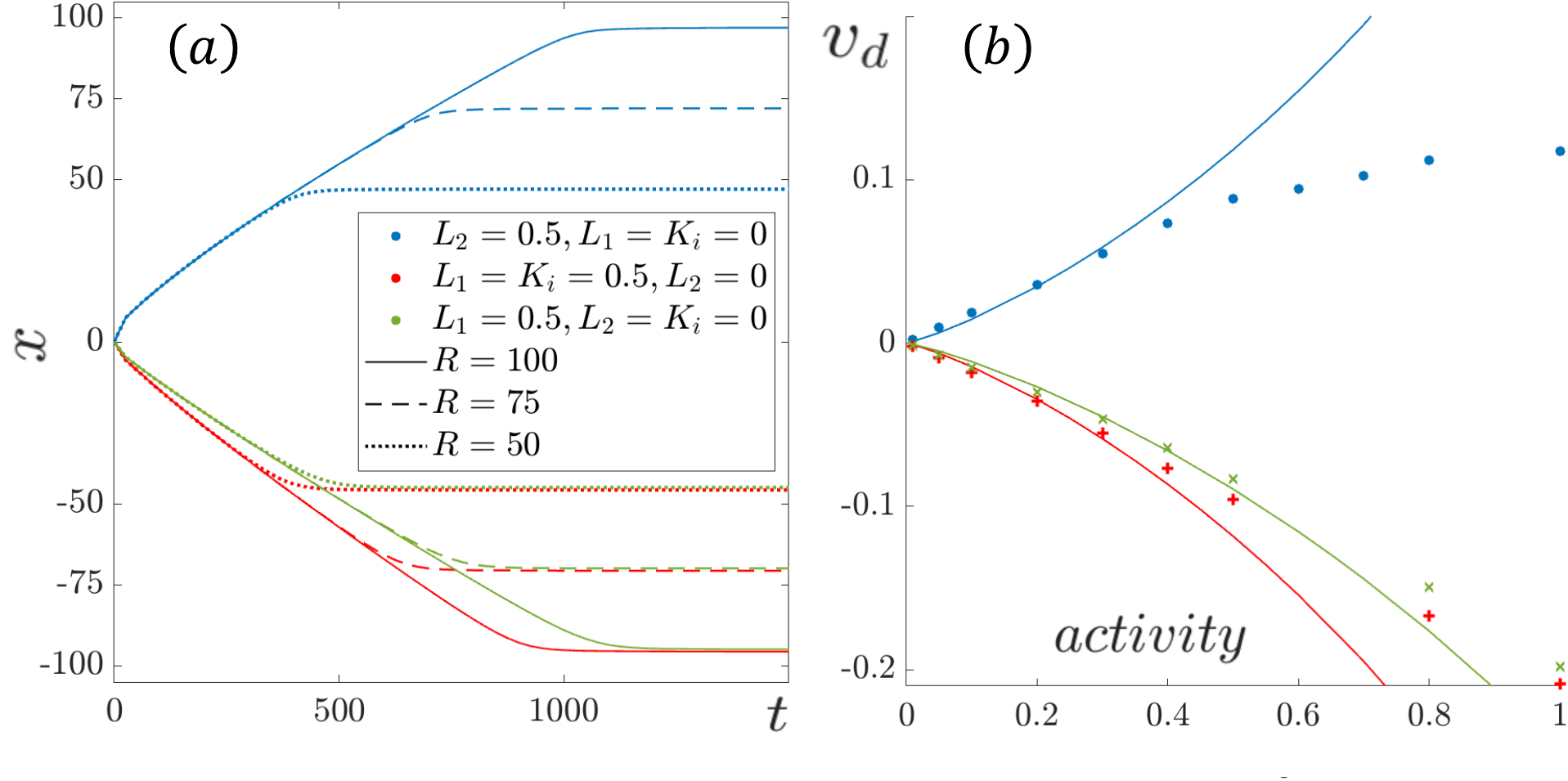}
		\par\end{centering}
	\caption{ Panel (a): Motion of the $+1/2$ defects in three active systems and three different system-sizes $R$, showing that defects always self-propel with a well-defined speed (i.e., independent of $R$), irrespectively of whether the active non-linearity $L_2$ is present. At late times, defects reach the boundary of the domain and stop. Only the defect with $L_2\neq0$ self-propels along its polarity while the other two do so anti-parallel to it. Here, $K_i\equiv K_1, K_2$.
	Panel (b): Defect self-propulsion speed $v_d$ in systems with various levels of activity, obtained by varying $L_2$ (blue), $K_1=K_2=L_1=\tilde{K}$ (red), and $L_1$ (green) and setting other coefficients to $0$, showing that $v_d$ has a comparable magnitude in all cases.
	 Dots are measurements from the simulations, lines analytical predictions at small activity obtained from eq. \eqref{eq:defect-speed-perturbative}, showing very good agreement, while eq. \eqref{eq:v_d-wrong} gives a wrong prediction at arbitrarily low activity.	
}
	\label{Fig2}
\end{figure}

In the following we compute the defect shape analytically by means of a perturbative expansion in the core and in the far field. These results are compared to numerical simulations of eq. \eqref{eq:minimal-nematics} performed with a finite-element solver implemented in MATLAB (see~\cite{supp} for details). These are performed in a circular domain of radius $R$. To induce a $\pm 1/2$ defect in the bulk we impose at the boundary the nematic order corresponding to a $\pm 1/2$ defect in a single Frank constant system infinitely far away from the core. This implementation is depicted in  Fig. \ref{Fig1}.

We start by recalling the shape of $\pm 1/2$ topological defects in the single Frank constant case. 
These are characterised by the amplitude of the nematic order obeying $S\simeq a_0 r$ and $S\simeq\sqrt{2}(1-1/(2r^2))$ respectively at small and large distances, while the angle field is given by $\theta=\pm \phi/2+\phi_0$ everywhere. Here, $r=\sqrt{x^2+y^2}$ is the distance from the position of the defect (that we consider to be at $r=0$), $a_0\simeq 0.82$ is a numerical constant~\cite{Neu}, and $\phi_0$ is an arbitrary angle fixing the orientation of the defect. For readability, we report our theoretical calculations for $\phi_0=0$; however, we also state the results for $\phi_0=\pi/2$, which is the case displayed in Fig. \ref{Fig1} and the one we consider numerically all through the Letter. 

Before presenting our main results, we show why taking into account shape modification due to activity is crucial. To do so, we employ a method popular in active matter literature \cite{ronning2023spontaneous, Angheluta2,de2024mesoscale,ronning2023precursory,jacques2023aging} and apply it to compute the speed $v_d$ of the $+1/2$ defect along the $\hat{x}$ direction from eq. (\ref{eq:minimal-nematics}). This method relies on a technique developed by Halperin and Mazenko \cite{Mazenko1,Mazenko2}, but crucially also assumes that the shape at the defect core is given by the one in single Frank constant passive systems. Our calculation extends the one of~\cite{Angheluta2}, that was performed in wet active nematics, to the dry case and is detailed in~\cite{supp}.
We obtain
\qq\label{eq:v_d-wrong}
v_d=-L_2 a_0\,,
\qqq
while  $v_d=L_2 a_0$ when $\phi_0=\pi/2$.

Eq. \eqref{eq:v_d-wrong} has several consequences. First it implies that, in active systems with $L_2=0$, $+1/2$ defects should not self-propel. Below, we show numerically that this conclusion is incorrect. Second, even more worrying is the fact that eq. \eqref{eq:v_d-wrong} predicts that the passive system with free energy in eq. \eqref{eq:free-energy-2-frank} should host $+1/2$ topological defects that self-propel. This conclusion is certainly incorrect: self-propulsion of defects is forbidden in equilibrium by energy conservation. 
Third, a different methodology \cite{pismen1999vortices}---which does not make any assumption on the defect shape---was also used in the literature \cite{Pismen, Suraj_def, PisSag,Suraj_PNAS} to compute the speed of $+1/2$ defects to leading order in small activity. As shown in~\cite{supp}, this perturbative technique applied to eq. \eqref{eq:minimal-nematics} gives
\qq\label{eq:defect-speed-perturbative}v_d\log\left(\frac{3.29}{v_d}\right)\simeq0.07(K_1+K_2)-0.78L_2+0.64L_1,
\qqq
in disagreement with eq. \eqref{eq:v_d-wrong} (for $\phi_0=\pi/2$, the right hand side acquires a global minus sign). Consistently, eq.~\eqref{eq:defect-speed-perturbative} predicts $v_d=0$ in the equilibrium limit $L_i=K_i$, $i=1,2$.

The reason for these inconsistencies is that eq. \eqref{eq:v_d-wrong}  assumes a wrong shape of the defect, as is often done in the literature. As we show in~\cite{supp}, for passive systems with two Frank constants defined by eq. \eqref{eq:free-energy-2-frank}, 
using the correct defect shape derived later in this Letter, yields $v_d=0$ for $+1/2$ defects within the Halperin-Mazenko formalism \cite{Mazenko2}, as it must in any equilibrium system.

We now show numerically that $+1/2$ defects in active systems always self-propel irrespective of whether $L_2\neq0$,  that eq. \eqref{eq:defect-speed-perturbative} is validated at small activity, and that eq. \eqref{eq:v_d-wrong} is incorrect as expected. We do so by performing numerical simulations of eq. \eqref{eq:minimal-nematics} with our finite-elements code~\cite{supp} initialising the system with the defect placed at the centre.  
We interpolate $S$ and identify its minimum with the defect position; tracking the defect location in time allows us to measure its speed. Fig. \ref{Fig2}  reports the defect position as a function of time in three active cases and, for each of them, for three system sizes. 
We always observe the defect self-propelling until it reaches the boundary and then stopping. This is starkly different from the case of passive systems where only a small drift induced by boundary effects, which decreases upon increasing the system-size, is observed~\cite{supp}. 
Furthermore, as shown in Fig. \ref{Fig2}(b), defects self-propel irrespective of whether $L_2\neq0$, and their velocity is very well captured by eq. \eqref{eq:defect-speed-perturbative} at small activity. Interestingly, we observe that $v_d$ saturates at large activity when $L_2\neq0$ while it does not in the other active cases. An analytical explanation of this fact is missing so far.

  \begin{figure}
	\begin{centering}
		\includegraphics[width=1.\columnwidth]{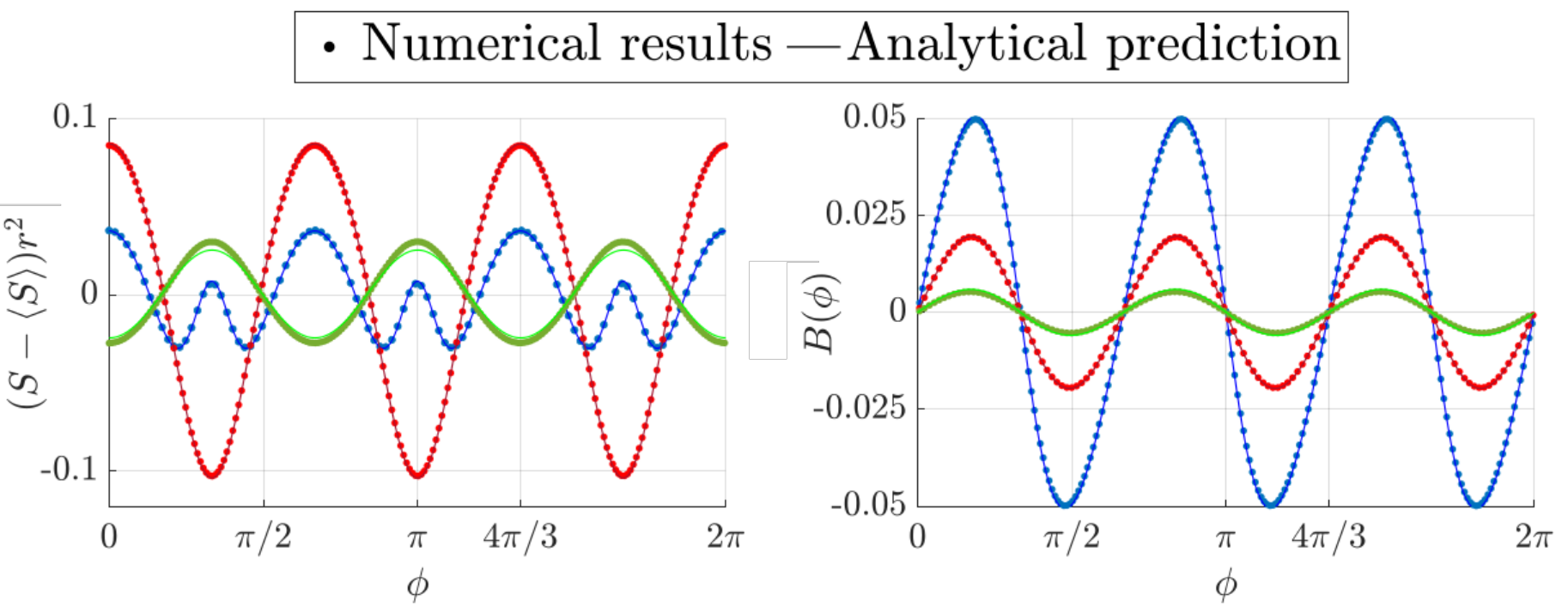}
		\par\end{centering}
	\caption{ Shape of the $-1/2$ topological defect in the far field encoded in the functions $A(\phi)$ and $B(\phi)$ of eq. (\ref{eq:generic_far}), (\ref{eq:generic_far-2}) found by solving eq. \eqref{eq:defect-12} and \eqref{eq:active-minus-half-B}. Analytical predictions (lines) and numerical simulations (dots) are in excellent agreement. In blue: passive system with two Frank constants ($K=1/2$); in red and green, respectively, active systems with $L_{2}=1/2, L_{1}=K_{1}=K_{2}=0$ and  $K_1=K_2=L_1=0.1, L_2=0$. The shape of the defects was measured at $r=15$ in domains with $R=100$ (blue and red) and $R=120$ (green).
 }\label{Fig3}
\end{figure}

We now discuss how the shape of defects is modified by activity, starting from the case of the $-1/2$. Inspired by the literature on passive liquid crystals~\cite{Neu, Hagan, Greenberg}, we make the following ansatz within the core 
 \qq\label{eq:generic_core}
S(r,\phi) &=& a(\phi) r^{m_s} +o( r^{m_s}),\\
\theta(r,\phi) &=&\pm \frac{\phi}{2} + b(\phi) r^{m_\theta}+ o(r^{m_\theta})\,,\label{eq:generic_core-2}
\qqq
and far away from it
\qq\label{eq:generic_far}
S(r,\phi) &=& \sqrt{2}+A(\phi) r^{-M_s} +o( r^{-M_s}),\\
\theta(r,\phi) &=& \pm\frac{\phi}{2} + B(\phi)+ O(r^{-M_\theta}),\,\label{eq:generic_far-2}
\qqq
where $m_s, m_\theta, M_s, M_\theta$ are strictly positive real numbers. 
Imposing that the defects have topological charge $(1/2\pi)\int_0^{2\pi} \partial_\phi\theta d\phi=\pm 1/2$ implies that $\int_0^{2\pi} a'(\phi)d\phi=\int_0^{2\pi} A'(\phi)d\phi=\int_0^{2\pi} b'(\phi)d\phi =\int_0^{2\pi} B'(\phi)d\phi = 0$,  and thus that $a(0)=a(2\pi)$, $A(0)=A(2\pi)$ and $b(0)=b(2\pi)$, $B(0)=B(2\pi)$. Our strategy is to insert eq. (\ref{eq:generic_core}-\ref{eq:generic_far-2}) in the static version of eq. \eqref{eq:minimal-nematics} and calculate all the unknown quantities in the two regimes $r\to0$ and $r\to\infty$. This is done by solving the resulting equations order by order in $r$ and in $1/r$.

The long but straightforward calculation is most  conveniently performed in polar coordinates and is described in~\cite{supp}. In all cases, we find $m_s=1$, $m_\theta=1$, $M_s=2$ and $a(\phi)= a_0$ constant. We further find $b(\phi)=0$ meaning that, even in active systems, the core field of $-1/2$ defects is the same as in the equilibrium single Frank constant liquid crystals. For the far-field, we obtain 
 \qq\label{eq:defect-12}
A(\phi )&=&\frac{\left[1-2 B'(\phi )\right]^2}{4}  \Big\{-2 \sqrt{2}\nonumber\\
&+&(K_2+2L_1) \cos [3 \phi -2 B(\phi )]\Big\}\,,
\qqq
while $B(\phi)$ solves
\begin{multline}\label{eq:active-minus-half-B}
8 B''(\phi )+4 \sqrt{2} \Big\{B'(\phi ) \sin [3 \phi -2 B(\phi )] \big[(K_2-2 L_2) B'(\phi )\\
-K_2+2 (L_1+L_2)\big]-L_1 B''(\phi ) \cos [3 \phi -2 B(\phi )]\Big\}\\
+\sqrt{2} \sin [3 \phi -2 B(\phi )] [K_2-2 (2 L_1+L_2)]=0\,,
\end{multline}
subject to the boundary condition $B(0)=B(2\pi)$.

Eq. \eqref{eq:defect-12} and \eqref{eq:active-minus-half-B} can be analytically solved perturbatively in the small activity limit. This gives
\begin{eqnarray}
	\label{farshape}
	B&=&\frac{(K_2-4L_1-2L_2)}{36\sqrt{2}}\sin 3\phi\,,\\
	A&=&-\frac{1}{\sqrt{2}}\left[1+\frac{(2L_1+4L_2-5K_2)}{6\sqrt{2}}\cos3\phi\right]\,.
\end{eqnarray}
In fact, the perturbative solution can be found even at higher orders in $K_{1,2}$ and $L_{1,2}$, but the resulting expressions are cumbersome and we do not report them here.
On the other hand, it is straightforward to solve eq. \eqref{eq:defect-12} and \eqref{eq:active-minus-half-B} numerically. 
This shows that the defect shape in active nematics is clearly distinct from the one found in passive liquid crystals with either one or two Frank constants (see Fig. \ref{Fig3}).

We now numerically test our analytical predictions for the shape of the $-1/2$ defect. We report the results for the equilibrium case with two Frank constants ($K=1/2$) and two active systems, although we observed similar agreement for other active cases we tested. 
In Fig.~\ref{Fig3} we plot the amplitude of the nematic order $S$ and the modification of the angle field $\theta$ with respect to the single Frank constant defect. To do so, we extract from simulations the functions $A$ and $B$ defined in eq. (\ref{eq:generic_far}), (\ref{eq:generic_far-2}) respectively by measuring $r^2(S-\langle S\rangle)$, where $\langle S\rangle$ is the angular average of $S$ at distance $r$ from the defect location, and $\theta+\phi/2$
sufficiently far both from the defect core and from the boundary. These simulations perfectly confirm  our analytical predictions obtained from eq. \eqref{eq:defect-12} and \eqref{eq:active-minus-half-B}. We also checked that similar agreement is obtained for the core field of the $-1/2$ defect (data not shown). 

  \begin{figure}
	\begin{centering}
		\includegraphics[width=1.\columnwidth]{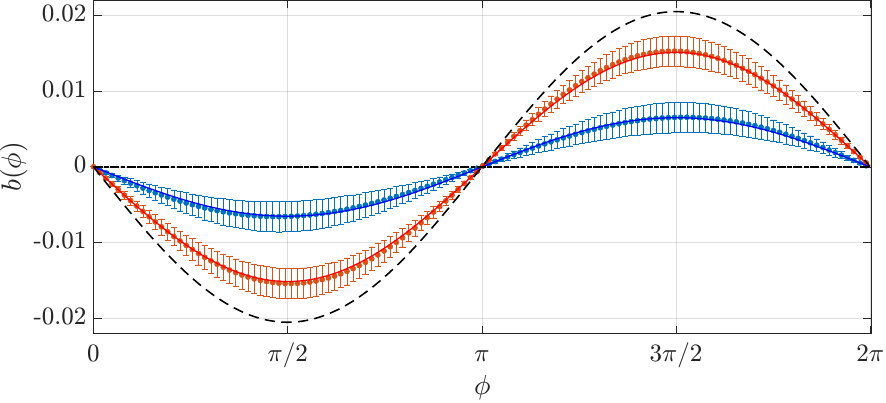}
		\par\end{centering}
	\caption{  Angular shape of the core of $+1/2$ defect as predicted theoretically (lines) by eq. (\ref{eq:vd-shape}) and found in our numerical simulations (points), showing perfect agreement. Red symbols correspond to $K_1=K_2=L_1=0, L_2=0.1$, blue ones to $K_1=K_2=L_1=0.1, L_2=0$, the dashed line is the defect shape for a two Frank constants defect with $K=0.1$ and the dashed-dotted line to the shape of a single Frank constant defect ($K=0$). Bars correspond to the error in measuring $b(\phi)$ coming from the uncertainty in the defect location due to the spatial discretisation employed in simulations. Other parameters: $R=10$, spatial discretisation $\Delta x_0=2\times 10^{-3}$ for $r<3$ and $\Delta x_0=0.5$ for $r>3$. $b(\phi)$ measured at $r=0.5$.
	}\label{Fig4}
\end{figure}

We now consider $+1/2$ defects. In the case of passive systems, their shape can be computed using exactly the same procedure as the one developed above for $-1/2$ defects~\cite{supp}, confirming earlier results~\cite{Dzyalo, Deem, Angheluta_SM, Schimming_Vinals,Thomas_hudson_PRL, Zhang_PNAS}. The case of active systems is however more delicate. We assume the system to be in the steady state, and hence the defect self-propels with a constant velocity ${\bf v}_d$.
Shifting to the co-moving reference frame ${\bf r}\to {\bf r}-{\bf v}_dt$, the defect must solve
\begin{equation}
	-({\bf v}_d\cdot \nabla) Q_{ij}=\mathcal{G}_{ij}\,.
\end{equation}
We then use the ansatz in eq. (\ref{eq:generic_core}-\ref{eq:generic_far-2}) and apply the same strategy as detailed above for $-1/2$ defects~\cite{supp}. We find that $m_s=m_\theta=1$, and $a(\phi)= a_0$ constant as in the single-Frank constant passive defect. Furthermore, the modification of the angle field near the core is found to be
\qq\label{eq:vd-shape}
b(\phi)=\frac{1}{4}(v_d+a_0L_2)\sin\phi\,.
\qqq
(For $\phi_0=\pi/2$, we obtain $b(\phi)=(1/4)(v_d-a_0L_2)\sin\phi$).
This replaces the predictions in eq. \eqref{eq:v_d-wrong} at arbitrarily large activity, and explicitly shows that the shape of the angle field in the core is directly linked to the self-propulsion speed. 

Next, we numerically verify our theoretical prediction in eq. (\ref{eq:vd-shape}). We measure both the self-propulsion speed $v_d$ and, going to the reference frame co-moving with the defect, the angle field modification encoded in $b(\phi)$. Fig. \ref{Fig4} shows that the numerically obtained shape perfectly matches the prediction in eq. (\ref{eq:vd-shape}): irrespective of whether $L_2\neq0$, the defect shape in active systems is clearly distinct from the one of in passive systems. 

While this Letter focused on single defect properties, our results imply that the shape of defects is also crucial for computing defect-defect interactions. Indeed, when defects are far apart, we can assume that they do not mutually modify their shape but rather that each of them imposes a boundary condition $\theta_{\text{ext}}$ for the other. Following standard literature~\cite{pismen1999vortices} we conclude that the interaction force on a defect of charge $s$ is given by $s^{-1}\bm{\epsilon}\cdot\nabla\theta_{\text{ext}}$ evaluated at the site of the defect, where $\bm{\epsilon}$ is the two-dimensional Levi-Civita tensor. Therefore, our results clearly imply that the interactions between defects depend on defect shape, and hence on activity; we however leave an examination of whether this implies that the interaction of topological defects can be non-reciprocal, as was argued to be the case in other active systems~\cite{AM_hex, Farzan_nem}, for future work. Notice that this is distinct from non-central forces between defects that arise even in equilibrium systems~\cite{Romano1, Romano2}. 

In conclusion, we considered a minimal dry model of active nematics, and showed that activity modifies the shape of defects. We have shown that neglecting such shape modifications can lead to incorrect conclusions even at arbitrarily small activity. 
For the $-1/2$ defect, we computed activity-induced shape modifications analytically both in the core and in the far field; for the $+1/2$ defect we unveiled a relation between the defect shape in the near field and the self-propulsion speed. 
Our analytics were perfectly matched by numerical simulations.

More generally our results can likely be extended to other active systems, for example to include fluid flow as is often done in describing active nematics~\cite{Doostmohammadi_rev, RMP, Raphael, Ano_2d, Aditi, AM_PNAS}, or to describe topological defects in active systems with other types of liquid-crystalline order. We show in \cite{supp} that modification of defect shape changes the self-propulsion speed of $+1/2$ defects even in wet systems.
We expect the methodology developed in this Letter to be the starting point to study further properties of topological defects in active systems, such as defect-defect interactions~\cite{sieberer2018topological, wachtel2016electrodynamic, sieberer2016lattice}, defect nucleation~\cite{cates2023classical,Suraj_def,giomi2013}, and their interplay with activity~\cite{Suraj_PNAS} or density gradients~\cite{wang2023patterning}. 

In this respect, the case of systems with hexatic order is of particular interest: it was shown experimentally~\cite{Palacci:12,van2019interrupted}, in large-scale simulations of self-propelled particles at high density~\cite{caporusso2022dynamics,caporusso2020micro,paliwal2020role}, as well as in biological tissues~\cite{pasupalak2020hexatic,Giomi_epithelia,tang2024cell}, that defects in active hexatics self-organise in defect lines or clusters. A natural question is whether this is induced by non-reciprocal defect interactions caused by activity-driven far-field defect shape changes. Another open question concerns interactions between defects in polar liquid crystals \cite{Angheluta_pol, Carles_pol1}. Recent numerical works suggested that an active polar phase at constant density is destroyed by the unbinding of defects~\cite{Besse_pol, Solon_XY, Popli}. A theoretical understanding of this is lacking and requires the calculation of defect interactions as well as the modification of the $-1$ defect shape. Analogous calculations may also prove useful in active systems that break translation symmetry such as smectics \cite{Toner_dis, pisegna2024emergent}, columnars \cite{kole2024chirality} and solids \cite{Vitelli_dis, Brauns}.

\begin{acknowledgments}
AM and CN  acknowledge the
support of the ANR grant PSAM. CN acknowledges the support of the INP-IRP grant IFAM and from the Simons Foundation. SB, AM and CN thank the Isaac Newton Institute for Mathematical Sciences for support and hospitality during the program ``Anti-diffusive dynamics: from sub-cellular to astrophysical scales'' during which a part of this work was undertaken. AM acknowledges a TALENT fellowship awarded by CY Cergy Paris Universit\'e. This research was supported in part by grants NSF PHY-2309135, PHY-1748958 and NSF PHY11-25915 and Gordon and Betty Moore Foundation Grant No. 2919 to the Kavli Institute for Theoretical Physics.
\end{acknowledgments}

\bibliographystyle{apsrev4-1}
\bibliography{ref}

\newpage
\newpage
\onecolumngrid

\appendix

\section{Equations of motion for the order parameter}
In this Appendix, we display the explicit form of the dry active nematic model in eq. (1) of the main text, referred thereafter as the equation of motion,
in Cartesian and polar coordinates, as well as in complex notation. We start by displaying eq. (1) of the main text here for convenience, in its non-dimensionalised form:
\begin{multline}
	\label{startEq}
\p_t Q_{ij}=
Q_{ij} -  \text{Tr}(\bsfQ^2)Q_{ij} + \nabla^2 Q_{ij} 
-\frac{K_1}{4} \mathcal{L}_{ij}\, \mathrm{Tr}(\bsfQ^2 )
+\frac{K_2}{2} Q_{kl} \mathcal{L}_{ij} Q_{kl}
+ L_1 Q_{kl}\nabla_k\nabla_l Q_{ij}
+L_2 (\nabla_k Q_{kl}) (\nabla_l Q_{ij})\,.
\end{multline}

\subsection{Equations of motion for $S$ and $\theta$ in Cartesian coordinates}
This form of the equations of motion is going to be useful in Secs. \ref{sec:linear-stability} and \ref{app:nematics}; therefore, we report it explicitly here. Inserting eq. (2) of the main text in \eqref{startEq} we obtain
\qq\label{eq:actCarttheta}
\p_t \theta &=&
 \nabla^2\theta +2 (\nabla\log S)\cdot(\nabla\theta)
+\frac{K_1}{8S}\Big[ \frac{1}{2}\sin(2\theta) (\p_x^2 S^2-\p_y^2 S^2)-\cos(2\theta) \p_x\p_y S^2 \Big]\nonumber
\\&+&\frac{K_2}{8}\Big\{ 
2\cos(2\theta)\Big[
\p_x\p_y S -4 S (\p_x \theta)(\p_y\theta)\Big]
-\sin(2\theta)\Big[\p_x^2S-\p_y^2S -4S ((\p_x\theta)^2-(\p_y\theta)^2)
\Big]
\Big\}
\\&+&L_1\Big\{
\cos(2\theta)\Big[
(\p_xS)(\p_x\theta)-(\p_yS)(\p_y\theta)
+\frac{S}{2}(\p_x^2\theta-\p_y^2\theta)
\Big]
+\sin(2\theta)\Big[
(\p_xS)(\p_y\theta)+(\p_yS)(\p_x\theta) + S\p_x\p_y\theta
\Big]
\Big\}\nonumber
\\&+&\frac{L_2}{2}\Big\{\cos(2\theta)\Big[
(\p_xS)(\p_x\theta)-(\p_yS)(\p_y\theta)+4S(\p_x\theta)(\p_y\theta)
\Big]\nonumber\\
&&\qquad\qquad+\sin(2\theta)\Big[
(\p_xS)(\p_y\theta)+(\p_yS)(\p_x\theta)
-2S((\p_x\theta)^2-(\p_y\theta)^2)
\Big]\Big\}\,.\nonumber
\qqq
and
\begin{multline}\label{eq:actCartS}
		\partial_{t}S = S\left(1-\dfrac{1}{2}S^{2}\right) + \nabla^{2}S + 4 S \left[ (\partial_{x}\theta)^{2} - (\partial_{y}\theta)^{2} \right]\\
		 +\dfrac{K_{1}}{4} \Big\{ \cos(2\theta) \left[ (\partial_{x}S)^{2} - (\partial_{y}S)^{2} - S (\partial_{x}^{2}S-\partial_{y}^{2}S)\right]
		 -2\sin(2\theta)[\partial_{x}S\partial_{y}S+S\partial_{x}\partial_yS] \Big\}
		 \\+ \dfrac{K_{2}}{4}S \Big\{ 2\sin(2\theta) \left[\partial_{x}\partial_yS - 4S\partial_{x}\theta\partial_{y}\theta\right]
	 + \cos(2\theta) \left[ \partial_{x}^{2}S-\partial_{y}^{2}S -4 S \left((\partial_{x}\theta)^{2} - (\partial_{y}\theta)^{2} \right)\right]\Big\}\\
		 + \dfrac{L_{1}}{2} S \Big\{ 2\sin(2\theta) \left[\partial_{x}\partial_yS + 4S\partial_{x}\theta\partial_{y}\theta\right]
	 + \cos(2\theta) \left[ \partial_{x}^{2}S-\partial_{y}^{2}S -4 S \left((\partial_{x}\theta)^{2} - (\partial_{y}\theta)^{2} \right)\right]\Big\}\\
		 + \dfrac{L_{2}}{2} \Big\{ 2\sin(2\theta) \left[\partial_{x}S\partial_{y}S + S(\partial_{y}S\partial_{y}\theta - \partial_{x}S\partial_{x}\theta)\right]
	 + \cos(2\theta) \left[ 2S(\partial_{x}S\partial_{y}\theta - \partial_{y}S\partial_{x}\theta) + (\partial_{x}S)^{2} - (\partial_{x}S)^{2} \right] \Big\}\,.
\end{multline}
In the single Frank constant passive limit, $K_1=K_2=L_1=L_2=0$. In this case, the equation for $\theta$ reduces to $\p_t \theta =\nabla^2\theta +2 (\nabla\log S)\cdot(\nabla\theta)$ and while the one for $S$ is given by the first line of \eqref{eq:actCartS}.

\subsection{Equations of motion for $S$ and $\theta$ in polar coordinates}
The form of eq. \eqref{startEq} in polar coordinates will be needed for computing the defects shape and we report it here explicitly for convenience. They read
\begin{multline}\label{eq:polActTheta}
		\partial_{t}\theta = \dfrac{1}{r^{2}S}\Big[ 2r^{2}\partial_{r}S\partial_{r}\theta + 2\partial_{\phi}S\partial_{\phi}\theta + S\left(\partial_{\phi}^{2}\theta + r\partial_{r}\theta + r^{2}\partial_{r}^{2}\theta\right)\Big]\\
	 +\dfrac{K_{1}}{8r^{2}S}\Big\{r^{2}\left(\partial_{r}S\right)^{2}\sin\left(2(\theta-\phi)\right) + S\Big[ r^{2}\partial_{r}^{2}S\sin\left(2(\theta-\phi)\right) 
	- \left(\partial_{\phi}^{2}S + r\partial_{r}S\right)\sin\left(2(\theta-\phi)\right) \\+ 2\left(\partial_{\phi}S-r\partial_{r}\partial_\phi S\right)\cos\left(2(\theta-\phi)\right)\Big]
		- \left(\partial_{\phi}S\right)^{2}\sin\left(2(\theta-\phi)\right) - 2r\partial_{r}S\partial_{\phi}S\cos\left(2(\theta-\phi)\right) \Big\}
		\\- \dfrac{K_{2}}{8r^{2}} \Big\{\left( 4S(\partial_{\phi}\theta)^{2}-\partial_{\phi}^{2}S -r\partial_{r}S\right)\sin\left(2(\theta-\phi)\right)
		+ r\Big[ r\partial_{r}^{2}S\sin\left(2(\theta-\phi)\right) - 2\partial_r\partial_\phi S\cos\left(2(\theta-\phi)\right) \\+ 4S\partial_{r}\theta
		\left( 2\partial_{\phi}\theta\cos\left(2(\theta-\phi)\right) - r\partial_{r}\theta\sin\left(2(\theta-\phi)\right) \right) + 2\partial_{\phi}S\cos\left(2(\theta-\phi)\right)\Big]\Big\}
		\\- \dfrac{L_{1}}{2r^{2}} \Big\{ S \Big[ -r^{2}\partial_{r}^{2}\theta\cos\left(2(\theta-\phi)\right) + 2\left( \partial_{\phi}\theta-r\partial_r\partial_\phi\theta\right)\sin\left(2(\theta-\phi)\right)
		+ \left(\partial_{\phi}^{2}\theta + r\partial_{r}\theta\right)\cos\left(2(\theta-\phi)\right)\Big]\\ - 2r\partial_{r}S\big[ \partial_{\phi}\theta\sin\left(2(\theta-\phi)\right)
		+ r\partial_{r}\theta\cos\left(2(\theta-\phi)\right)\big] + 2\partial_{\phi}S\big[ \partial_{\phi}\theta\cos\left(2(\theta-\phi)\right) - r\partial_{r}\theta\sin\left(2(\theta-\phi)\right) \big]\Big\}
		\\+\dfrac{L_{2}}{2r^{2}} \Big\{ -2r^{2}S\left(\partial_{r}\theta\right)^{2}\sin\left(2(\theta-\phi)\right) + r\partial_{r}\theta \Big[ \big( r\partial_{r}S
		+ 4S\partial_{\phi}\theta\big)\cos\left(2(\theta-\phi)\right) + \partial_{\phi}S\sin\left(2(\theta-\phi)\right)\Big]
	\\	- \partial_{\phi}\theta \Big[-\big(r\partial_{r}S + 2S\partial_{\phi}\theta\big)\sin\left(2(\theta-\phi)\right) + \partial_{\phi}S\cos\left(2(\theta-\phi)\right)\Big]\Big\}\,.
\end{multline}
and
\begin{multline}\label{eq:polActS}
		\partial_{t}S=  S - \dfrac{1}{2}S^{3}+ \dfrac{1}{r^{2}} \Big\{ -4S \left[ r^{2}(\partial_{r}\theta)^{2} + (\partial_{\phi}\theta)^{2}\right]+\partial_{\phi}^{2}S + r\partial_{r}S + r^{2}\partial_{r}^{2}S\Big\} \\
		 + \dfrac{K_{1}}{4r^{2}} \Big\{ -r^{2}(\partial_{r}S)^{2}\cos\left(2(\theta-\phi)\right) + S \Big[ -r^{2}\partial_{r}^{2}S\cos\left(2(\theta-\phi)\right) 
		 + 2(\partial_{\phi}S-r\partial_r\partial_\phi S)\sin\left(2(\theta-\phi)\right) + (\partial_{\phi}^{2}S + r\partial_{r}S)\cos\left(2(\theta-\phi)\right) \Big]\\
		- 2r\partial_{r}S\partial_{\phi}S\sin\left(2(\theta-\phi)\right) + (\partial_{\phi}S)^{2}\cos\left(2(\theta-\phi)\right)\Big\} \\
		+\dfrac{K_{2}}{4r^{2}} S \Big\{ \Big[-\partial_{\phi}^{2}S - r\partial_{r}S + 4S(\partial_{\phi}\theta)^{2}\Big]\cos\left(2(\theta-\phi)\right)
		+r \Big[ 2\partial_r\partial_\phi S\sin\left(2(\theta-\phi)\right) + r\partial_{r}^{2}S\cos\left(2(\theta-\phi)\right) \\
		 -4S\partial_{r}\theta \left( r\partial_{r}\theta\cos\left(2(\theta-\phi)\right) + 2\partial_{\phi}\theta\sin\left(2(\theta-\phi)\right)\right)\Big]
		- 2\partial_{\phi}S\sin\left(2(\theta-\phi)\right)\Big\}\\
		+ \dfrac{L_{1}}{4r^{2}} S \Big\{ \Big[-\partial_{\phi}^{2}S - r\partial_{r}S + 4S(\partial_{\phi}\theta)^{2}\Big]\cos\left(2(\theta-\phi)\right)
		+r \Big[ 2\partial_r\partial_\phi S\sin\left(2(\theta-\phi)\right) + r\partial_{r}^{2}S\cos\left(2(\theta-\phi)\right) 
		 \\-4S\partial_{r}\theta \left( r\partial_{r}\theta\cos\left(2(\theta-\phi)\right) + 2\partial_{\phi}\theta\sin\left(2(\theta-\phi)\right)\right)\Big]
		 - 2\partial_{\phi}S\sin\left(2(\theta-\phi)\right)\Big\}\\
		 + \dfrac{L_{2}}{2r^{2}} \Big\{ 2\partial_{\phi}S \Big[rS\partial_{r}\theta\cos\left(2(\theta-\phi)\right)
		 + \Big( r\partial_{r}S + S\partial_{\phi}\theta \Big) \sin\left(2(\theta-\phi)\right)\Big]\\
	+ r\partial_{r}S \big[ \left( r\partial_{r}S+2S\partial_{\phi}\theta \right)\cos\left(2(\theta-\phi)\right) 
		 -2rS\partial_{r}\theta\sin\left(2(\theta-\phi)\right) \big] - (\partial_{\phi}S)^{2}\cos\left(2(\theta-\phi)\right) 
		\Big\}
\end{multline}

In the two Frank constant passive limit, $K_1=K_2=L_1=L_2=K$, and these equations read
\begin{multline}\label{eq:polPassTheta}
		\partial_{t}\theta=\dfrac{1}{r^{2}S}\Big[ 2r^{2}\partial_{r}S\partial_{r}\theta + 2\partial_{\phi}S\partial_{\phi}\theta + S\left(\partial_{\phi}^{2}\theta + r\partial_{r}\theta + r^{2}\partial_{r}^{2}\theta\right)\Big]\\+ \dfrac{K}{8r^{2}S}\Big(2\cos2(\phi-\theta)[6S(r^2\partial_rS\partial_r\theta-\partial_\phi S\partial_\phi\theta)-r\partial_\phi S\partial_rS-2S^2(\partial_\phi^2\theta-r^2\partial_r^2\theta+r\partial_r\theta-2r\partial_r\theta\partial_\phi\theta)]\\-\sin2(\phi-\theta)\{12rS(\partial_\phi S\partial_r\theta+\partial_\phi\theta\partial_r S)+r^2(\partial_r S)^2-(\partial_\phi S)^2+4S^2[(\partial_\phi\theta)^2-2\partial_\phi\theta+2r\partial_r\partial_\phi\theta-r^2(\partial_r\theta)^2]\}
		\Big)\\
\end{multline}
\begin{multline}\label{eq:polPassS}
	\partial_{t}S=  S - \dfrac{1}{2}S^{3}+ \dfrac{1}{r^{2}} \Big\{ -4S \left[ r^{2}(\partial_{r}\theta)^{2} + (\partial_{\phi}\theta)^{2}\right]+\partial_{\phi}^{2}S + r\partial_{r}S + r^{2}\partial_{r}^{2}S\Big\} \\
	+\frac{K}{4r^2}\Big(\cos2(\phi-\theta)\{r^2(\partial_rS)^2-(\partial_\phi S)^2+4rS(\partial_\phi S\partial_r\theta+\partial_\phi\theta\partial_r S)+12S^2[(\partial_\phi\theta)^2-r^2(\partial_r\theta)^2]-2S(\partial_\phi^2S+r\partial_\phi S-r^2\partial_r^2S)\}\\-2\sin 2(\phi-\theta)[r\partial_rS\partial_\phi S-12rS^2\partial_r\theta\partial_\phi\theta+2S(\partial_\phi\theta\partial_\phi S-\partial_\phi S-r^2\partial_rS\partial_r\theta+r\partial_\phi\partial_r S)]
	\Big)
\end{multline}

\subsection{Equations of motion for the nematic order parameter in complex representation}
\label{CompEq}
In Sec. \ref{sec:Mazenko} it will be convenient to express eq. \eqref{startEq} in complex notation ($*$ denotes the complex conjugation below). We do so here by introducing the complex order parameter $\Psi(z,z^*)$
\qq
\Psi=Q_{xx}+iQ_{xy}=(S/2)(\cos2\theta+i\sin2\theta)=(S/2)e^{2i\theta}
\qqq
as well as the partial derivative with respect to $z=x+iy$ which is given by $\partial=(1/2)[\partial_x-i\partial_y]$. Eq. \eqref{startEq}  reads
\qq
	\label{Psiac}
	\partial_t\Psi&=&(1-2|\Psi|^2)\Psi+4\partial\partial^*\Psi-K_1\partial^*(\Psi^*{\partial^*}\Psi+\Psi\partial^*\Psi^*)+K_2(\Psi{\partial^*}^2\Psi^*+\Psi^*{\partial^*}^2\Psi)\nonumber \\
	&+&2L_1(\Psi^*{\partial^*}^2\Psi+\Psi\partial^2\Psi)+{ 2L_2(\partial^*\Psi^*\partial^*\Psi+(\partial\Psi)^2)}\,.
\qqq

We note that 
under rotation of the coordinate by an angle $\psi$ we have $\theta\to \theta-\psi$, implying 
\qq
&\Psi\to\Psi e^{-2i\psi}\\
&x\to x\cos\psi+y\sin\psi\\
&y\to -x\sin\psi+y\cos\psi\,.
\qqq
Since $\partial_x=\cos\psi\partial_x+\sin\psi\partial_y$, $\partial_y=-\sin\psi\partial_x+\cos\psi\partial_y$, we have
\qq
\partial\to e^{i\psi}\partial\,.
\qqq
These transformations can be used to immediately check that our equation is invariant under a joint rotation of space and $\Psi$, as it should be.

In the two Frank constant limit, this reduces to
\begin{equation}
	\label{PassEqcomplex}
	\partial_t\Psi=(1-2|\Psi|^2)\Psi+4\partial\partial^*\Psi+K[2\Psi^*{\partial^*}^2\Psi+\partial^2(\Psi^2)].
\end{equation}

\section{Linear stability}\label{sec:linear-stability}
It can be immediately checked that the linear stability of the disordered state is unaffected by activity. We consider here the linear stability of eq. \eqref{startEq} around a perfectly ordered state identified by $S=\sqrt{2}$ and $\theta=0$. From \eqref{eq:actCarttheta} and \eqref{eq:actCartS}, we get
\begin{equation}
\p_t \delta\theta =\nabla^2\delta\theta +\frac{L_1}{\sqrt{2}} (\p_x^2-\p_y^2)\delta\theta+\frac{K_2-K_1}{4}\p_x\p_y \delta S\,,
\end{equation}
\begin{equation}
\p_t \delta S = -2\delta S +\frac{\sqrt{2}}{4}(K_1-K_2-2L_1) (\p_x^2-\p_y^2)\delta S +  \nabla^2 \delta S\,,
\end{equation}
for small perturbations $\delta S$ and $\delta\theta$ around this state. Computing the eigenfrequencies from the spatiotemporally Fourier-transformed versions of these equations, we find that the ordered state is stable at all wavenumbers when

\begin{equation}\label{eq:linear-stability}
 |L_1|<\sqrt{2}\,\,\,\,;\,\,\,\
 | K_1-K_2-2L_1 |<\sqrt{8}\,.
\end{equation}
Notice that these two conditions in \eqref{eq:linear-stability} reduce to a single one in the passive limit of a two Frank constant system. When performing numerical simulations, we made sure that eq. (\ref{eq:linear-stability}) are satisfied.

\section{Spin wave theory for $\theta$ for fluctuations about a defect-free ordered state with $S=S_0$ }
In this section we display the hydrodynamic spin wave theory for the angle field of our model for small fluctuations about a nematic state ordered along the $\hat{x}$ axis. This reads
\begin{equation}\label{eq:actCarttheta1}
	\p_t \theta =
	\nabla^2\theta +\frac{(K_2-2L_2)S_0}{2}\Big\{ 
	(\sin2\theta)[(\p_x\theta)^2-(\p_y\theta)^2]-2\cos(2\theta) \p_x \theta\p_y\theta
	\Big\}
	+\frac{L_1S_0}{2}\Big[
	(\cos2\theta)(\p_x^2\theta-\p_y^2\theta)
	+2\sin(2\theta)\p_x\p_y\theta
	\Big]
	\,,
\end{equation}
which contains arbitrary powers of the angle field. When expanded to $O(\theta^2)$, this reads
\begin{equation}\label{eq:actCarttheta2}
	\p_t \theta =\left(1+\frac{L_1S_0}{2}\right)\partial_x^2\theta+\left(1-\frac{L_1S_0}{2}\right)\partial_y^2\theta+2L_1S_0\theta\partial_x\partial_y\theta+(2L_2-K_2)S_0\partial_x\theta\partial_y\theta
	\,.
\end{equation}
A stochastic version of \eqref{eq:actCarttheta2} was examined in \cite{mishra2010dynamic} using perturbative renormalisation group methods (see their Eq. 8 whose non-stochastic part is the same as \eqref{eq:actCarttheta2} with the identification $1+L_1S_0/2\to A_1$, $1-L_1S_0/2\to A_2$, $2L_1S_0\to\lambda_2$ and $(2L_2-K_2)S_0\to\lambda_1$). This identification also explicitly satisfies the relation $2(A_1-A_2)=\lambda_2$ that they identify as being required by rotation invariance.

\section{Calculation of defect velocity in active systems using a single Frank constant defect shape}
In this section, we calculate the velocity ${\bf v}_d$ of a $+1/2$ defect in active systems by assuming that it retains the shape it would have had in a single Frank constant passive nematic. This approximation is often made in active materials (see, for instance, \cite{Angheluta2}) and the calculation of the self-propulsion uses a method originally due to Mazenko \cite{Mazenko2, Mazenko1}. Here we adapt it to our dry active nematics. In the complex notation of Eq. \eqref{PassEqcomplex}, and defining a complex defect velocity $v_d^c=v_{d_x}+iv_{d_y}$, we have from \cite{Mazenko2, Mazenko1, Angheluta2} that
\begin{equation}
	\label{velMaz}
	v_d^c=\left[\frac{\partial^*\Psi\partial_t\Psi^*-\partial^*\Psi^*\partial_t\Psi}{\partial\Psi\partial^*\Psi^*-\partial\Psi^*\partial^*\Psi}\right]_{z=0}\,.
\end{equation}
The defect in a single Frank constant system solves the equations
\begin{equation}
	\label{thetaeq}
	S\partial\partial^*\theta+\partial^*S\partial\theta+\partial S\partial^*\theta=0\,
\end{equation}
and
\begin{equation}
	\label{Seq}
	\left(1-\frac{1}{2}S^2\right)S-16\partial\theta\partial^*\theta+4\partial\partial^*S=0\,,
\end{equation}
which are obtained from \eqref{Psiac} with $K_1=K_2=L_1=L_2=0$, using the definition of $\Psi$.
The angle field has the solution
\begin{equation}
	\theta=\frac{i}{2}s\log\left[\frac{z^*}{z}\right]\,,
\end{equation}
where $s$ is the topological charge of the defect.
This, when plugged into \eqref{Seq}, yields
\begin{equation}
	\left(1-\frac{1}{2}S^2\right)S-\frac{S}{zz^*}+4\partial\partial^*S=0\,,
\end{equation}
implying that for $z,z^*\to 0$, $S\sim a_0\sqrt{zz^*}$ where $a_0$ is a numerical constant.
This implies
\begin{equation}
	\label{Rel1}
	\lim_{z,z^*\to 0}\partial\Psi=\frac{a_0(1+2s)}{4}\lim_{z,z^*\to 0}\left(\frac{z}{z^*}\right)^{s-\frac{1}{2}}= \frac{a_0}{2} \,\, \text{for}\, s=+\frac{1}{2}\,,
\end{equation}
and is $0$ when $s=-1/2$. Similarly,
\begin{equation}
	\label{Rel2}
	\lim_{z,z^*\to 0}\partial^*\Psi=\frac{a_0(1-2s)}{4}\lim_{z,z^*\to 0}\left(\frac{z}{z^*}\right)^{s+\frac{1}{2}}=\frac{a_0}{2} \,\, \text{for}\, s=-\frac{1}{2}\,,
\end{equation}
and $0$ when $s=+1/2$. With these results, the defect velocity can be obtained as
\begin{equation}
	v_d^c=-\frac{2}{a_0}\partial_t\Psi(z,z^*=0)\,,
\end{equation}
for a $+1/2$ defect and
\begin{equation}
	v_d^c=\frac{2}{a_0}\partial_t\Psi^*(z,z^*=0)\,,
\end{equation}
for a $-1/2$ defect.
We note that $\partial\partial^*\Psi=0$ for both $\pm1/2$ defects. We now use Eq. \eqref{Psiac} for the dynamics of $\Psi$.
Eliminating all terms that are obviously $0$ at the site of defects (where $\Psi=0$)
\begin{equation}
	\partial_t\Psi|_{z=0}=[2L_2(\partial\Psi)^2]_{z=0}\,.
\end{equation}
Therefore, from \eqref{Rel1} and \eqref{Rel2} we find that the velocity of a $+1/2$ defect is
\begin{equation}
	v_d^c=-L_2a_0.
\end{equation}
Since $(\partial^*\Psi^*)^2=0$ for a $-1/2$ defect, $v_d^c=0$ in this case. This implies that ${\bf v}_d=-L_2a_0\hat{{\bf x}}$ for an active $+1/2$ defect.

It is now easy to see using \eqref{Psiac} why only $L_2$ contributes to this velocity: the terms with the coefficients $K_2$ and $L_1$ vanish at the site of the defect because $\Psi=\Psi^*=0$ there. Expanding the term $K_1$, we either have terms that vanish at the site of the defect because $\Psi=\Psi^*=0$  or have the term $\partial^*\Psi^*\partial^*\Psi$. At the site of a $-1/2$ defect, $\partial^*\Psi^*=0$ by \eqref{Rel1} and at the site of a $+1/2$ defect, $\partial^*\Psi=0$ by \eqref{Rel2} implying that $\partial^*\Psi^*\partial^*\Psi=0$ at the site of $\pm1/2$ defects and that $K_1$ also doesn't contribute to the defect velocity if the shape of the defect is constrained to be the one that it would have in single Frank constant equilibrium system.

\section{Self-propulsion $+1/2$ defect assuming unperturbed defect shape -- do even defects in passive system self-propel?}\label{sec:Mazenko}
As we discussed earlier, the dynamical equation in \eqref{startEq} reduces to that of an equilibrium two Frank constant nematic with a model-A-like dynamics \cite{chaikin,hohen} in the limit $K_1=K_2=L_1=L_2=K$. The defect velocity we found in the previous section is puzzling in the light of this: it doesn't vanish in the passive limit but becomes ${\bf v}_d=-Ka_0\hat{{\bf x}}$. This is not an artefact of the method used to calculate the defect velocity; instead, it is due to assuming a defect shape that is not the one that minimises the free energy in a two Frank constant system. To establish this, we now calculate the velocity of defects in a passive two Frank constant system described by equations 
\eqref{eq:polPassTheta} and \eqref{eq:polPassS} when their shapes are taken to be solutions of those equations with $K=0$ using a different method.

\subsection{Defect velocity calculated from Eqs. \eqref{eq:polPassTheta} and \eqref{eq:polPassS}}
\label{app:passmov}
To start with, we consider a topologically non-trivial solution of the equations
\begin{equation}
	\label{eq:1FCtheta}
0 = \dfrac{1}{r^{2}S}\Big[ 2r^{2}\partial_{r}S\partial_{r}\theta + 2\partial_{\phi}S\partial_{\phi}\theta + S\left(\partial_{\phi}^{2}\theta + r\partial_{r}\theta + r^{2}\partial_{r}^{2}\theta\right)\Big]\,,
\end{equation}
\begin{equation}\label{eq:1FCS}
	0=  S - \dfrac{1}{2}S^{3}+ \dfrac{1}{r^{2}} \Big\{ -4S \left[ r^{2}(\partial_{r}\theta)^{2} + (\partial_{\phi}\theta)^{2}\right]+\partial_{\phi}^{2}S + r\partial_{r}S + r^{2}\partial_{r}^{2}S\Big\}\,,
\end{equation}
in a punctured plane. Assuming that $S$ depends only on $r$, we see that \eqref{eq:1FCtheta} has the well-known solution $\theta=\pm\phi/2$ for a defect with topological charge $\pm1/2$ with one of the principal axes along $\hat{\bf x}$.
Plugging this in \eqref{eq:1FCS}, we get
\begin{equation}\label{eq:1FCSfin}
	0=  S - \dfrac{1}{2}S^{3}+ \dfrac{1}{r^{2}}  (-S  + r\partial_{r}S + r^{2}\partial_{r}^{2}S)\,,
\end{equation}
which, near the core of the defect at $r=0$ has the solution
\begin{equation}
	\label{Score1FC}
	S=a_0r+o(r)\,.
\end{equation}
We now consider the possibility that when this defect shape is used for a system with two Frank constants, the solution is a defect that moves with a constant velocity ${\bf v}_d$ which, in polar coordinates, has the components $(v_{d_r},v_{d_\phi})$. In a frame moving with the velocity ${\bf v}_d$, the defect then solves the equation
\begin{multline}\label{eq:polPassThetamot}
	-\frac{v_{d_\phi}\partial_\phi\theta}{r}-v_{d_r}\partial_r\theta=\dfrac{1}{r^{2}S}\Big[ 2r^{2}\partial_{r}S\partial_{r}\theta + 2\partial_{\phi}S\partial_{\phi}\theta + S\left(\partial_{\phi}^{2}\theta + r\partial_{r}\theta + r^{2}\partial_{r}^{2}\theta\right)\Big]\\+ \dfrac{K}{4r^{2}S}\Big(2\cos2(\phi-\theta)[6S(r^2\partial_rS\partial_r\theta-\partial_\phi S\partial_\phi\theta)-r\partial_\phi S\partial_rS-2S^2(\partial_\phi^2\theta-r^2\partial_r^2\theta+r\partial_r\theta-2r\partial_r\theta\partial_\phi\theta)]\\-\sin2(\phi-\theta)\{12rS(\partial_\phi S\partial_r\theta+\partial_\phi\theta\partial_r S)+r^2(\partial_r S)^2-(\partial_\phi S)^2+4S^2[(\partial_\phi\theta)^2-2\partial_\phi\theta+2r\partial_r\theta\partial_\phi\theta-r^2(\partial_r\theta)^2]\}
	\Big)\,,
\end{multline}
\begin{multline}\label{eq:polPassSmot}
		-\frac{v_{d_\phi}\partial_\phi S}{r}-v_{d_r}\partial_r S=  S - \dfrac{1}{2}S^{3}+ \dfrac{1}{r^{2}} \Big\{ -4S \left[ r^{2}(\partial_{r}\theta)^{2} + (\partial_{\phi}\theta)^{2}\right]+\partial_{\phi}^{2}S + r\partial_{r}S + r^{2}\partial_{r}^{2}S\Big\} \\
	+\frac{K}{2r^2}\Big(\cos2(\phi-\theta)\{r^2(\partial_rS)^2-(\partial_\phi S)^2+4rS(\partial_\phi S\partial_r\theta+\partial_\phi\theta\partial_r S)+12S^2[(\partial_\phi\theta)^2-r^2(\partial_r\theta)^2]-2S(\partial_\phi^2S+r\partial_\phi S-r^2\partial_r^2S)\}\\-2\sin 2(\phi-\theta)[r\partial_rS\partial_\phi S-12rS^2\partial_r\theta\partial_\phi\theta+2S(\partial_\phi\theta\partial_\phi S-\partial_\phi S-r^2\partial_rS\partial_r\theta+r\partial_\phi\partial_r S)]
	\Big)\,.
\end{multline}
Using the fact that in the single Frank constant limit, $S$ only depends on $r$ and $\theta$ only on $\phi$, we find from \eqref{eq:polPassThetamot}
\begin{equation}
	v_{d_\phi}=K\sin2(\phi-\theta)\Bigg[\frac{S(\partial_\phi\theta)^2}{r}-\frac{2S\partial_\phi\theta}{r}+{3\partial_\phi\theta\partial_rS}+\frac{r(\partial_rS)^2}{4S}\Bigg]\,.
\end{equation}
Now using the single Frank constant solution for $\theta=\pm \phi/2$ and $S$ (see eq. \eqref{Score1FC}) for a single Frank constant defect, 
we find that $v_\phi=0$ for $-1/2$ defects and $v_\phi=Ka_0\sin\phi$ for a $+1/2$ defect. Similarly, from \eqref{eq:polPassSmot}, we get 
\begin{equation}
	v_{d_r}=\frac{K\cos 2(\phi-\theta)}{a_0}\Bigg[\frac{\partial_rS^2}{4r}-\frac{3S^2(\partial_\phi\theta)^2}{r^2}-\frac{\partial_rS^2\partial_\phi\theta}{2r}-\frac{(\partial_rS)^2}{4}\Bigg].
\end{equation}
which gives $v_r=0$ for a $-1/2$ defect and $v_r=-Ka_0\cos\phi$ for a $+1/2$ defect. Noting that a vector ${\bf u}=(u_x,0)$ in Cartesian coordinates has the components $u_r=u_x\cos\phi$ and $u_\phi=-u_x\sin\phi$, we obtain the defect velocity ${\bf v}_d=-Ka_0\hat{{\bf x}}$.

This result is obviously incorrect: a defect in the absence of any external forcing cannot move perpetually in an equilibrium system. This calculation presents a reductio ad absurdum argument against using a single Frank constant defect shape in attempting to calculate its velocity.


\section{Perturbative calculation of defect speed in dry active nematics}
We start by writing the dynamics of the complex order parameter $\Psi(z,z^*)$ in terms of real space variables
\begin{multline}
	\label{EOMPsiXY}
	\partial_t\Psi=\Psi-2|\Psi|^2\Psi+\nabla^2\Psi+\frac{L_1}{2}\left[\Psi(\partial_x^2\Psi-2i\partial_x\partial_y\Psi-\partial_y^2\Psi)+\Psi^*(\partial_x^2\Psi+2i\partial_x\partial_y\Psi-\partial_y^2\Psi)\right]\\+\frac{L_2}{2}\left[(\partial_x\Psi)^2-(\partial_y\Psi)^2-\partial_y\Psi\partial_y\Psi^*-2i\partial_y\Psi\partial_x\Psi+i(\partial_y\Psi\partial_x\Psi^*+\partial_x\Psi\partial_y\Psi^*)+\partial_x\Psi\partial_x\Psi^*\right]\\+\frac{K_1}{4}\left[2(\partial_y\Psi-i\partial_x\Psi)(\partial_y\Psi^*-i\partial_x\Psi^*)+\Psi^*(\partial_y^2\Psi-2\partial_x\partial_y\Psi-\partial_x^2\Psi)+\Psi(\partial_y^2\Psi^*-2\partial_x\partial_y\Psi^*-\partial_x^2\Psi^*)\right]\\+\frac{K_2}{4}\left[\Psi^*(\partial_x^2\Psi+2i\partial_x\partial_y\Psi-\partial_y^2\Psi)+\Psi(\partial_x^2\Psi^*+2i\partial_x\partial_y\Psi^*-\partial_y^2\Psi^*)\right]\,,
\end{multline}
while that for the two Frank constant model is 
\begin{equation}
	\partial_t\Psi=\Psi-2|\Psi|^2\Psi+\nabla^2\Psi+\frac{K}{2}\left[-(\partial_y\Psi+i\partial_x\Psi)^2+\Psi(\partial_x^2\Psi-2i\partial_x\partial_y\Psi-\partial_y^2\Psi)+\Psi^*(\partial_x^2\Psi+2i\partial_x\partial_y\Psi-\partial_y^2\Psi)\right]\,.
\end{equation}

We now use a standard method of matched asymptotic expansion \cite{pismen1990, pismen1999vortices} to calculate the speed of $+1/2$ defects in the limit in which $K_1,K_2,L_1,L_2\ll 1$. This method was used by \cite{Pismen}---and, following that work, by \cite{Suraj_def}---to calculate the defect speed in active systems. We extend it to cover the dry case considered in this work and show that eq. (5) of the main text holds. As discussed in the main text, our results show that in the equilibrium limit $K_1=K_2=L_1=L_2=K$, the defect speed vanishes. 

Transforming to a frame co-moving with the defect, we have the equation
\begin{multline}
	\label{EOMPsiXYv}
	0={\bf v}_d\cdot\nabla\Psi+\Psi-2|\Psi|^2\Psi+\nabla^2\Psi+\frac{L_1}{2}\left[\Psi(\partial_x^2\Psi-2i\partial_x\partial_y\Psi-\partial_y^2\Psi)+\Psi^*(\partial_x^2\Psi+2i\partial_x\partial_y\Psi-\partial_y^2\Psi)\right]\\+\frac{L_2}{2}\left[(\partial_x\Psi)^2-(\partial_y\Psi)^2-\partial_y\Psi\partial_y\Psi^*-2i\partial_y\Psi\partial_x\Psi+i(\partial_y\Psi\partial_x\Psi^*+\partial_x\Psi\partial_y\Psi^*)+\partial_x\Psi\partial_x\Psi^*\right]\\+\frac{K_1}{4}\left[2(\partial_y\Psi-i\partial_x\Psi)(\partial_y\Psi^*-i\partial_x\Psi^*)+\Psi^*(\partial_y^2\Psi-2\partial_x\partial_y\Psi-\partial_x^2\Psi)+\Psi(\partial_y^2\Psi^*-2\partial_x\partial_y\Psi^*-\partial_x^2\Psi^*)\right]\\+\frac{K_2}{4}\left[\Psi^*(\partial_x^2\Psi+2i\partial_x\partial_y\Psi-\partial_y^2\Psi)+\Psi(\partial_x^2\Psi^*+2i\partial_x\partial_y\Psi^*-\partial_y^2\Psi^*)\right]\,.
\end{multline}

Following \cite{Pismen, pismen1999vortices}, we now rescale ${\bf v}_d\to\varepsilon{\bf v}_d$, $K_i\to\varepsilon K_i$, $L_i\to\varepsilon L_i$, where $\varepsilon\ll 1$ is a bookkeeping parameter for the perturbation theory. We further expand $\Psi=\Psi_0+\varepsilon\Psi_1$. From \eqref{EOMPsiXYv}, we obtain the zeroth order in $\varepsilon$ equation for $\Psi$:
\begin{equation}
	\label{0ord}
\Psi_0-2|\Psi_0|^2\Psi_0+\nabla^2\Psi_0=0\,.
\end{equation}
The solution for this is \cite{pismen1999vortices}
\begin{equation}
	\Psi_0(r,\phi)=\frac{r}{\sqrt{2}}\sqrt{\frac{0.34+0.07r^2}{1+0.41r^2+0.07r^4}}e^{\pm i\phi}\equiv\frac{\tilde{S}_0(r)}{\sqrt{2}}e^{\pm i\phi}\,,
\end{equation}
for $\pm1/2$ defects. The $O(\epsilon)$ equation is 
\begin{equation}
	\label{1ord}
	\mathcal{H}(\Psi_1,\Psi_1^*)+\mathcal{I}=0
\end{equation}
where
\begin{equation}
	\label{linop}
	\mathcal{H}(\Psi_1,\Psi_1^*)\equiv \Psi_1-2\Psi_0^2\Psi_1^*-4|\Psi_0|^2\Psi_1+\nabla^2\Psi_1\,,
\end{equation}
\begin{multline}
\mathcal{I}\equiv{\bf v}_d\cdot\nabla\Psi_0+\frac{L_1}{2}\left[\Psi_0(\partial_x^2\Psi_0-2i\partial_x\partial_y\Psi_0-\partial_y^2\Psi_0)+\Psi_0^*(\partial_x^2\Psi_0+2i\partial_x\partial_y\Psi_0-\partial_y^2\Psi_0)\right]\\+\frac{L_2}{2}\left[(\partial_x\Psi_0)^2-(\partial_y\Psi_0)^2-\partial_y\Psi_0\partial_y\Psi_0^*-2i\partial_y\Psi_0\partial_x\Psi_0+i(\partial_y\Psi_0\partial_x\Psi_0^*+\partial_x\Psi_0\partial_y\Psi_0^*)+\partial_x\Psi_0\partial_x\Psi_0^*\right]\\+\frac{K_1}{4}\left[2(\partial_y\Psi_0-i\partial_x\Psi_0)(\partial_y\Psi_0^*-i\partial_x\Psi_0^*)+\Psi_0^*(\partial_y^2\Psi_0-2\partial_x\partial_y\Psi_0-\partial_x^2\Psi_0)+\Psi_0(\partial_y^2\Psi_0^*-2\partial_x\partial_y\Psi_0^*-\partial_x^2\Psi_0^*)\right]\\+\frac{K_2}{4}\left[\Psi_0^*(\partial_x^2\Psi_0+2i\partial_x\partial_y\Psi_0-\partial_y^2\Psi_0+\Psi_0(\partial_x^2\Psi_0^*+2i\partial_x\partial_y\Psi_0^*-\partial_y^2\Psi_0^*)\right]\,.
\end{multline}
The linear operator in \eqref{linop} has a pair of eigenfunctions with zero eigenvalues given by $\nabla\Psi_0$. This can be seen by taking the gradient of \eqref{0ord} which yields
\begin{equation}
\mathcal{H}(\nabla\Psi_0,\nabla\Psi_0^*)\equiv\nabla\Psi_0-4|\Psi_0|^2\nabla\Psi_0-2\Psi_0^2\nabla\Psi_0^*+\nabla^2\nabla\Psi_0=0\,.
\end{equation}
Since the defect is not expected to rotate in an achiral system by symmetry, we do not consider another eigenfunction with a zero eigenvalue that exists \cite{Pismen}.

Now since the homogeneous equation corresponding to \eqref{1ord} has a nontrivial solution, Eq. \eqref{1ord} is only solvable if $\mathcal{I}$ is orthogonal to the eigenfunction with zero eigenvalue i.e., to $\nabla\Psi_0$ \cite{pismen1999vortices}. As shown in \cite{pismen1999vortices}, applying the just-discussed Fredholm alternative to \eqref{1ord} in the infinite plane is inconsistent; instead, it should be applied in a circle of radius $r_0$ large compared to the core but small compared to $1\ll r_0\ll\varepsilon^{-1}$ (note that $r_0$ and all quantities are dimensionless; returning to dimensionalised form, we see that $\kappa/|{\bf v}_d|$ is a lengthscale. This is the large lengthscale in the model).

Applying the solvability condition to \eqref{1ord} in the aforementioned domain, we obtain
\begin{equation}
	\label{Fred}
	\mathrm{Re}\left[\int_0^{r_0}rdr\int_0^{2\pi}\nabla\Psi_0^*\mathcal{I}(r,\phi)d\phi+r_0\int_0^{2\pi}\left(\nabla\Psi_0^*\partial_r\Psi_1-\Psi_1\partial_r\nabla\Psi_0^*\right)_{r=r_0}d\phi\right]=0\,.
\end{equation}
We can explicitly evaluate the first term in this expression using the solution of $\Psi_0$ that corresponds to a $+1/2$ defect. The integrals corresponding to the terms $K_1, K_2, L_1, L_2$ converge at large $r_0$ and, for these integrals, the upper limit can be taken to infinity \cite{Pismen}. In practice, we compute these integrals numerically by setting a large $r_0$. Defining $\tilde{\mathcal{I}}=\mathcal{I}-{\bf v}_d\cdot\nabla\Psi_0$, we obtain
\begin{equation}
	\label{activexp}
\mathrm{Re}\left[\int_0^{r_0}rdr\int_0^{2\pi}\nabla\Psi_0^*\tilde{\mathcal{I}}(r,\phi)d\phi\right]=\left(1.236L_2-1.014L_1-0.111K_2-0.111K_1\right)\hat{x}\,.
\end{equation}
Note that, as required, this vanishes when $L_1=L_2=K_1=K_2=K$. Note further that terms that vanish when $\Psi=0$ nevertheless contribute to the speed; this fact was already implicitly recognised by \cite{Pismen, Suraj_def}.

The integral containing ${\bf v}_d$ in \eqref{Fred} diverges at large $r_0$. Noting that for a $+1/2$ defect, 
\begin{equation}
	\nabla\Psi_0=e^{i\phi}\left[\partial_r\tilde{S}_0(\cos\phi,\sin\phi)-\frac{i\tilde{S}_0}{r}(\sin\phi,-\cos\phi)\right]\,,
\end{equation}
we obtain
\begin{equation}
	\label{vdexp}
	\mathrm{Re}\left[\int_0^{r_0}rdr\int_0^{2\pi}\nabla\Psi_0^*{\bf v}_d\cdot\nabla\Psi_0 d\phi\right]=\frac{\pi}{2}{\bf v}_d\int_0^{r_0}\left[\frac{\tilde{S}_0^2}{r}+r\left(\partial_r\tilde{S}_0\right)^2\right]dr=\frac{\pi}{2}{\bf v}_d\log\frac{r_0}{1.126}\,.
\end{equation}

We now need to calculate the angular integral in \eqref{Fred} involving $\Psi_1$, and do so extending to the present case an argument developed in~\cite{Pismen}. Since, at large $r_0$, we have $\tilde{S}_0\to 1-1/2r_0^2$, and $\tilde{S}_0=1-O(\varepsilon)$ when $r_0\sim O(1/\sqrt{\varepsilon})$. We will assume in the following that $\Psi_1$ also decays as $1/r_0^2$ or faster (and it is shown below this assumption to be self-consistent for the angle field defining $\Psi_1$)~\cite{Pismen,pismen1999vortices}. Further, active terms are of the form $\nabla\Psi\nabla\Psi$ or $\Psi\nabla\nabla\Psi$, i.e., have two powers of the field and two powers of gradients; these terms decay faster than $1/r_0^2$ at large $r_0$; indeed this was the fact that allowed us to obtain \eqref{activexp} by extending the upper limit of the integral to infinity. Moreover, these terms are multiplied by an $O(\varepsilon)$ quantity. Therefore, at large $r_0$ (i.e., $r_0\sim O(1/\sqrt{\varepsilon})$), they are smaller than either the advective term (which has one fewer power of gradient and field) or the usual elastic term (which is not multiplied by a small quantity). This implies that at large scales, the shape modification of the defects is dominated by the \emph{motion} of the defect (irrespective of the cause; in passive systems, this is typically due to smooth phase gradients due to other defects; for $+1/2$ defects in active nematics that we consider here, this is due to the self-propulsion, ultimately due to the active nonlinear terms). Therefore, for $r_0\sim O(1/\sqrt{\varepsilon})$ or larger, the solution for $\Psi_1$ to leading order in $\varepsilon$ can be obtained by solving for the phase $\vartheta$ of $\Psi_1$ alone via the equation

\begin{equation}
	\label{phsEq}
	{\bf v}_d\cdot\nabla\vartheta+\nabla^2\vartheta=0\,.
\end{equation}
Eq. \eqref{phsEq} implies a solution \cite{Pismen, Radzihovsky,Denniston, pismen1999vortices, pismen1990, Ryskin, Suraj_def}
	\begin{equation}
		\label{solshp}
		\nabla\vartheta=\pm\frac{1}{4}e^{-{\bf v}_d\cdot{\bf r}/2}\bm{\epsilon}\cdot\left[{\bf v}_dK_0\left(\frac{|{\bf v}_d|r}{2}\right) -|{\bf v}_d|\hat{{\bf r}}K_1\left(\frac{|{\bf v}_d|r}{2}\right)\right]\,,
	\end{equation}
	where $\bm{\epsilon}$ is the two-dimensional Levi-Civita tensor and $K_0(x)$ and $K_1(x)$ are Bessel functions of the first kind.
Using this solution at $r_0$ (where $r_0\sim O(1/\sqrt{\varepsilon})$ such that this solution can be applied)
and following the steps discussed in \cite{Pismen,pismen1999vortices}, we obtain
\begin{equation}
	\label{ffieldexp}
	\mathrm{Re}\left[r_0\int_0^{2\pi}\left(\nabla\Psi_0^*\partial_r\Psi_1-\Psi_1\partial_r\nabla\Psi_0^*\right)_{r=r_0}d\phi\right]=-\frac{\pi}{2}{\bf v}_d\log\left(\frac{|{\bf v}_d|r_0 }{4}e^{\gamma_E-1/2}\right)\,,
\end{equation}
where we have not retained a term related to an external phase gradient (which we assume to vanish) since we are interested in calculating the self-propulsion of the defect in the absence of any such gradient \cite{Pismen, Suraj_def, pismen1999vortices}. Here, $\gamma_E\approx 0.577$ is the Euler constant \cite{Pismen, pismen1999vortices}.

We now reassemble \eqref{Fred} using \eqref{activexp}, \eqref{vdexp} and \eqref{ffieldexp} to obtain
\begin{equation}
\frac{\pi}{2}{\bf v}_d\log\left(\frac{4 }{1.126|{\bf v}_d|}e^{1/2-\gamma_E}\right)+\left(1.236L_2-1.014L_1-0.111K_2-0.111K_1\right)\hat{x}=0\,.
\end{equation}
Since $|{\bf v}_d|\sim O(\varepsilon)\ll1$, the logarithm is positive. The defect velocity can be expressed as
\begin{equation}
	\label{finspd}
{\bf v}_d\log\left(\frac{3.29}{|{\bf v}_d|}\right)=-\frac{2}{\pi}\left(1.236L_2-1.014L_1-0.111K_2-0.111K_1\right)\hat{x}=-\left(0.787L_2-0.645L_1-0.071K_2-0.071K_1\right)\hat{x}\,.
\end{equation}
The opposite sign with respect to the numerics reported in the main text is due to the opposite orientation of the defect in the numerics which are initialised with $\theta=\phi/2+\pi/2$. Note that we can solve for the value of ${\bf v}_d$ from this equation and replace it in \eqref{solshp} to obtain an expression for the shape of the far field of the $+1/2$ defect, in terms of the parameters of our model, to leading order in $K_i$ and $L_i$.

 \section{Shape of $-1/2$ defects}
  We now describe our calculation of the shape of the $-1/2$ defect which does not self-propel even in active systems. We start with the static version of eq. \eqref{startEq} in polar coordinates i.e., \eqref{eq:polActTheta} and \eqref{eq:polActS} with the left-hand-side set to $0$.

 \subsection{$-1/2$ defect shape in the far field}
 We start by examining the shape of $-1/2$ defect in the $r\to\infty$ limit. As explained in the main text, we start from the ansatz
 \qq\label{eq:equilibrium-act-Ansatz+half-defect-far}
 S(r,\phi) &=& \sqrt{2}+A(\phi) r^{-M_s} +o( r^{-M_s})\\
 \theta(r,\phi) &=& \frac{\phi}{2} + B(\phi)+ O(r^{-M_\theta})\,
 \qqq
 where $\int_0^{2\pi} A'(\phi)d\phi =\int_0^{2\pi} B'(\phi)d\phi = 0$,  implying $A(0)=A(2\pi)$ and $B(0)=B(2\pi)$.
 
 We first plug this ansatz into Eq. \eqref{eq:polActS} for $S$ (with $\partial_tS=0$). The coefficient of the term that decays the slowest as $r\to\infty$ has to vanish. From this, we get the condition
 \begin{equation}
 	r^{-M_{s}} A + \dfrac{1}{r^{2}}\left(-\dfrac{1}{2} + B'\right)^{2}\Big[ 2\sqrt{2} - (K_2+2L_1)\cos\left(3\phi-2B\right) \Big] = 0\,.
 \end{equation}
 This implies $M_s=2$ and
 \qq 
 A=\frac{\left(1-2 B'\right)^2}{4}  \left[-2 \sqrt{2}+(K_2+2L_1) \cos (3 \phi -2 B)\right]\,.
 \qqq
We then consider the equation for $\theta$, \eqref{eq:polActTheta}, with $\partial_t\theta=0$. The slowest decaying term as $r\to\infty$ again goes as $1/r^2$. Requiring its coefficient to vanish implies the following equation for $B(\phi)$:
 \begin{equation}\label{eq:active-minus-half-B}
 	8 B''+4 \sqrt{2} \Big\{B'\sin (3 \phi -2 B) \big[(K_2-2 L_2) B'
 	-K_2+2 (L_1+L_2)\big]-L_1 B''\cos (3 \phi -2 B)\Big\}
 	+\sqrt{2} \sin (3 \phi -2 B) [K_2-2 (2 L_1+L_2)]=0\,.
 \end{equation}
with the boundary conditions $B(0)=B(2\pi)=B_0$. 
Eq. (\ref{eq:active-minus-half-B}) can be solved numerically for arbitrary $B_0$. We can also solve this perturbatively for $K_1,K_2,L_1,L_2\ll1$. To do this, we expand $B=\varepsilon B_1+\varepsilon^2 B_2...$, $A=-1/\sqrt{2}(1+\varepsilon A_1+\varepsilon^2 A_2...)$, where $\varepsilon$ is a bookkeeping parameter we perform the perturbation in. We take $K_1\sim K_2\sim L_1\sim L_2\sim \varepsilon$. Inserting this in \eqref{eq:active-minus-half-B}, we get to first order in $\varepsilon$,
\begin{equation}
	8\varepsilon B_1''+\sqrt{2}[K_2-2(2L_1+L_2)]\sin3\phi=0\,.
\end{equation}
This has the solution
\begin{equation}
	\varepsilon B_1=\frac{(K_2-4L_1-2L_2)}{36\sqrt{2}}\sin 3\phi\,.
\end{equation}
The $O(\varepsilon)$ equation for $A_1$ is 
\begin{equation}
	(K_2+2L_1)\cos3\phi+2\sqrt{2}\varepsilon(A_1+4B')=0\,,
\end{equation}
which has the solution
\begin{equation}
\varepsilon A_1=\frac{2L_1+4L_2-5K_2}{6\sqrt{2}}\cos3\phi\,.
\end{equation}
To $O(\varepsilon^2)$, the solutions of $B$ and $A$ are
\begin{equation}
	\label{eq:act-halfBshape}
	B=\frac{(K_2-4L_1-2L_2)}{36\sqrt{2}}\sin 3\phi-\frac{(7K_2-34L_1-14L_2)[K_2-2(2L_1+L_2)]}{10368}\sin6\phi\,,
\end{equation}
\begin{equation}
		\label{eq:act-halfAshape}
	A=-\frac{1}{\sqrt{2}}\left\{1+\frac{(2L_1+4L_2-5K_2)}{6\sqrt{2}}\cos3\phi+\frac{[K_2-2(2L_1+L_2)][18(K_2+L_1)-6L_2+(31K_2-4L_1-20L_2)\cos6\phi]}{432}\right\}\,.
\end{equation}
We can carry out the perturbation theory to higher order in $\varepsilon$.

Note that the shape of a passive two Frank constant defect may be obtained from \eqref{eq:act-halfBshape} and \eqref{eq:act-halfAshape} by setting $K_1=K_2=L_1=L_2=K$. This yields [to $O(\varepsilon^2)\equiv O(K^2)$]
\begin{equation}
	\label{-halfpassiveshape}
	B=-\frac{5K}{36\sqrt{2}}\sin 3\phi-\frac{205K^2}{10368}\sin6\phi\,,
\end{equation}
\begin{equation}
	A=-\frac{1}{\sqrt{2}}\left[1+\frac{K\cos3\phi}{6\sqrt{2}}-\frac{5K^2}{432}(7\cos6\phi+30)\right]\,.
\end{equation}

 \subsection{$-1/2$ defect shape near the core}\label{sub:-half-core}
We now calculate the shape of the $-1/2$ defect near the core. 
We use the ansatz that the shape of the defect is given by
\qq\label{eq:equilibrium-Kneq0-Ansatz-half-defect-core}
S(r,\phi) &=& a(\phi) r^{m_s} +o( r^{m_s})\\\nonumber
\theta(r,\phi) &=& -\frac{\phi}{2} + b(\phi) r^{m_\theta}+ o(r^{m_\theta})\,,
\qqq
as $r\to 0$.
Here, $m_s>0$ because $S$ vanishes at the core, $m_\theta>0$ in order to have a well defined defect at $r=0$, and $\int_0^{2\pi} b'(\phi)d\phi =\int_0^{2\pi} a'(\phi)d\phi= 0$, which implies $b(0)=b(2\pi)$ and $a(0)=a(2\pi)$.
As in the earlier sections, we inject this ansatz first in \eqref{eq:polPassS}, set $\partial_tS=0$ and examine the terms that diverge the fastest in the $r\to 0$ limit. Setting these to $0$, we get the equation
\qq
r^{m_s-2} \Big[
-a(\phi) +m_s^2a(\phi)+a''(\phi)
\Big]=0
\qqq
which readily gives
\qq
a(\phi)
=
a_{1}e^{\alpha \phi }+ a_{2}e^{-\alpha \phi }\,.
\qqq
where $\alpha=\sqrt{1-m_s^2}$.
Imposing now that $a(\phi)=a(2\pi+\phi)$ and that $a(\phi)$ is real and non-negative for all $\phi$, we conclude that 
\qq
a(\phi)=a_0
\qqq
is independent of $\phi$, and $m_s=1$. 

Carrying out the same analysis for the next fastest diverging terms, we get
\qq\label{eq:+half-core-intermediate-1}
4 a_0  r^{m_\theta-1} { b'(\phi )}=0\,.
\qqq
This implies that $m_\theta=1$ and $b(\phi)=b_0$ has to be independent of $\phi$ (since it must be periodic).
Injecting this result into \eqref{eq:polPassTheta} with $\partial_t\theta=0$, we obtain
\begin{equation}
	3 b_{0}m_{\theta} r^{m_{\theta} - 2} = 0\,,
\end{equation}
implying that $b_0=0$. 
Therefore, the shape of the core of the $-1/2$ defect in an active system is identical to the one in a single Frank constant system.

 \section{Shape of $+1/2$ defects in passive systems with two Frank constants}
The shape of $+1/2$ defects in {passive} two Frank constant systems may be obtained using a similar procedure since they do not self-propel. We present the details of this calculation in this section.
 Our starting point is the static version of eq. \eqref{startEq} in polar coordinates which, for the passive system is given by \eqref{eq:polPassTheta} and \eqref{eq:polPassS} with the left-hand-side set to $0$. 
 
 \subsection{$+1/2$ defect shape in the far field}
 We now study the $+1/2$ defect shape in the $r\to\infty$ limit. As explained in the main text, we start from the Ansatz
  \qq\label{eq:equilibrium-Kneq0-Ansatz+half-defect-far}
S(r,\phi) &=& \sqrt{2}+A(\phi) r^{-M_s} +o( r^{-M_s})\,,\\
\theta(r,\phi) &=& \frac{\phi}{2} + B(\phi)+ O(r^{-M_\theta})\,,
\qqq
where $\int_0^{2\pi} A'(\phi)d\phi =\int_0^{2\pi} B'(\phi)d\phi = 0$,  implying $A(0)=A(2\pi)$ and $B(0)=B(2\pi)$.

\begin{figure}[h]\center
		\centering
		\includegraphics[width=0.5\linewidth]{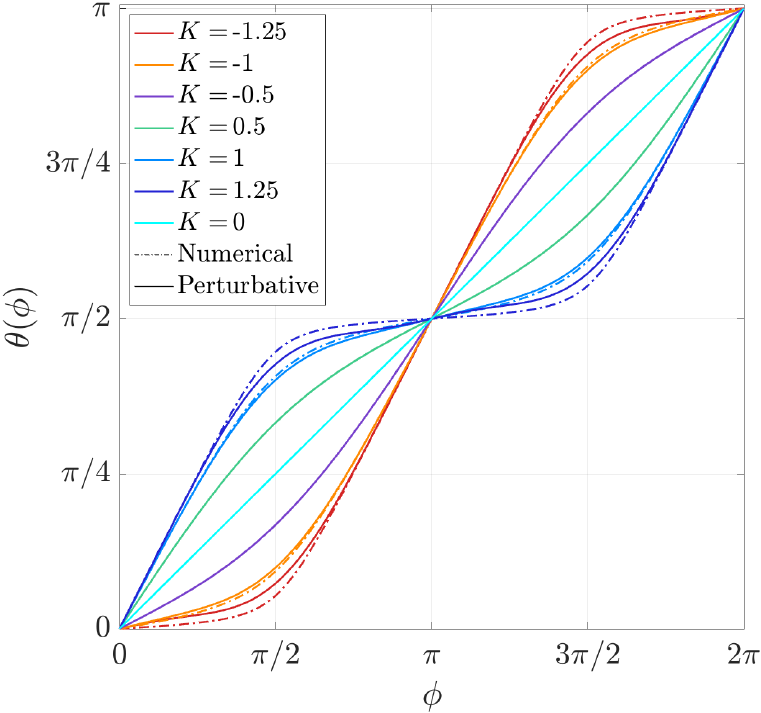}
		\caption{Far field shape of the angle field for passive $+1/2$ defect. The perturbative solution at order four discussed in eq. \eqref{app:eq-perturbative-sol} is indistinguishable from the one obtained by numerical solution of the ODE for $B(\phi)$ up to $K\approx 1$.}
		\label{Figpluhalf}
	\end{figure}

We first plug this ansatz into Eq. \eqref{eq:polPassS} for $S$ (with $\partial_tS=0$). The coefficient of the term that decays the slowest as $r\to\infty$ has to vanish. From this, we get the condition
\begin{equation}
	-r^{-M_{s}} A+ \dfrac{1}{r^{2}}\left(\dfrac{1}{2} + B'\right)^{2}\Big[ -2\sqrt{2} + 3K\cos\left(\phi-2B\right) \Big] = 0\,.
\end{equation}
Since $B$ has to be periodic and $K\neq 0$, this equation implies that $M_s=2$ and that 
\qq
A=\frac{\left(1+2 B'\right)^2}{4}  \left[-2 \sqrt{2}+3 {K} \cos (\phi -2 B)\right]\,.
\qqq
We then consider the equation for $\theta$, \eqref{eq:polPassTheta}, with $\partial_t\theta=0$. The slowest decaying term as $r\to\infty$ again goes as $1/r^2$. Requiring its coefficient to vanish implies the following equation for $B(\phi)$:
\begin{equation}
	\label{Beqplushalf}
K \left( \dfrac{3}{4} {+} B' -B'^{2}\right) \sin ( \phi -2 B)
= B''\left[ K \cos ( \phi -2 B)-\sqrt{2}\right]
\end{equation}
with the boundary conditions $B(0)=B(2\pi)=B_0$. Solutions for arbitrary $B_0$ can be found numerically. We can also obtain the solution for $K\ll 1$ perturbatively in $K$ for $B_0=0$. To do this, we first note that to zeroth order in $K$, $B(\phi)$ has to vanish since $B''=0$ has no non-constant periodic solution. In this case, $A=-1/\sqrt{2}$. We therefore expand $A$ and $B$ perturbatively in $K$ as $B=K B_1+K^2 B_2...$, $A=-1/\sqrt{2}(1+K A_1+K^2 A_2...)$. Inserting this in \eqref{Beqplushalf}, we get, to first order in $K$, 
\begin{equation}
	8B_1''+3\sqrt{2}\sin\phi=0\,.
\end{equation}
This has the solution
\begin{equation}
	B_1=\frac{3\sin\phi}{4\sqrt{2}}\,.
\end{equation}
The ${O}(K)$ equation for $A_1$ is
\begin{equation}
	3\cos\phi+2\sqrt{2}(A_1-4B_1')=0\,.
\end{equation}
This yields the solution
\begin{equation}
	A_1=\frac{3\cos\phi}{2\sqrt{2}}\,.
\end{equation}
Therefore, for small $K$, 
\begin{equation}
	B=\frac{3K\sin\phi}{4\sqrt{2}}\,,
\end{equation}
\begin{equation}
	A=-\frac{1}{\sqrt{2}}\left(1+\frac{3K\cos\phi}{2\sqrt{2}}\right)\,,
\end{equation}
to ${O}(K)$. To ${O}(K^2)$, the solutions are
\begin{equation}
	\label{+halfpassiveshape}
B=\frac{3K\sin\phi}{4\sqrt{2}}+\frac{3K^2}{128}\sin2\phi
\end{equation}
\begin{equation}
	A=-\frac{1}{\sqrt{2}}\left[1+\frac{3K\cos\phi}{2\sqrt{2}}+\frac{3K^2}{16}(\cos2\phi-6)\right]\,.
\end{equation}
We can carry out the perturbation theory to higher orders in $K$.
Doing this for $B(\phi)$ up to ${O}(K^4)$ yields
	\begin{equation}
		\theta=\frac{\phi}{2}+\frac{3K}{4\sqrt{2}}\sin\phi+\frac{3K^2}{128}\sin2\phi-\frac{K^3}{512\sqrt{2}}(51\sin\phi+11\sin3\phi)-\frac{K^4}{32768}(112\sin2\phi+19\sin4\phi)\,.\label{app:eq-perturbative-sol}
	\end{equation}
	We plot in Fig. \ref{Figpluhalf} both this solution that is perturbative in the second Frank constant and solution of the shape obtained by numerically integrating \eqref{Beqplushalf}, which is valid for arbitrary values of $K$. Our solutions closely match those obtained in \cite{Thomas_hudson_PRL, Zhang_PNAS}. 
	
	We further point out that Ref. \cite{Thomas_hudson_PRL} discusses the Dzyaloshinskii  solution \cite{Dzyalo} for a two Frank constant system. 
	Our Eq. \eqref{+halfpassiveshape} perturbative solution in passive nematics matches with the ones in \cite{Deem, Angheluta_SM} to first order in $K$ (i.e., the order up to which this expansion was carried out in those works); It has been shown in \cite{Angheluta_SM} that this solution is consistent with Dzyaloshinskii's \cite{Dzyalo}.

 \subsection{$+1/2$ defect shape near the core}\label{sub:+half-core}
 We now calculate the shape of the $+1/2$ defect near the core. 
We use the ansatz that the shape of the defect is given by
 \qq\label{eq:equilibrium-Kneq0-Ansatz+half-defect-core}
S(r,\phi) &=& a(\phi) r^{m_s} +o( r^{m_s})\\\nonumber
\theta(r,\phi) &=& \frac{\phi}{2} + b(\phi) r^{m_\theta}+ o(r^{m_\theta})\,,
\qqq
as $r\to 0$.
Here, $m_s>0$ because $S$ vanishes at the core, $m_\theta>0$ in order to have a well defined defect at $r=0$, and $\int_0^{2\pi} b'(\phi)d\phi =\int_0^{2\pi} a'(\phi)d\phi= 0$, which implies $b(0)=b(2\pi)$ and $a(0)=a(2\pi)$.
As in the earlier sections, we inject this ansatz first in \eqref{eq:polPassS}, set $\partial_tS=0$ and examine the terms that diverge the fastest in the $r\to 0$ limit. Setting these to $0$, we get the equation
\qq
r^{m_s-2}  \Big[
-a(\phi) +m_s^2a(\phi)+a''(\phi)
\Big]=0
\qqq
which readily gives
\qq
a(\phi)
=
a_{1}e^{\alpha \phi }+ a_{2}e^{-\alpha \phi }\,.
\qqq
where $\alpha=\sqrt{1-m_s^2}$.
Imposing now that $a(\phi)=a(2\pi+\phi)$ and that $a(\phi)$ is real and non-negative for all $\phi$, we conclude that 
\qq
a(\phi)=a_0
\qqq
is independent of $\phi$, and $m_s=1$. 

Carrying out the same analysis for the next fastest diverging terms, we get
 \qq\label{eq:+half-core-intermediate-2}
 a_0^2 K \cos\phi -4 a_0  r^{m_\theta-1} { b'(\phi )}=0\,.
 \qqq
This implies that $m_\theta=1$ and 
\qq
b(\phi)= b_{1} +\frac{a_0 K}{4} \sin\phi
\qqq
We now plug these results in the equation for $\theta$, \eqref{eq:polPassTheta} with $\partial_t\theta=0$, and demanding that the coefficient of the most diverging term vanishes, we get $b_1=0$. Thus, in the passive limit, we find that the $+1/2$ defect has the following shape near the core:
 \qq
S(r,\phi) &=& a_0r+o(r)\,,\\
\theta(r,\phi) &=& \frac{\phi}{2} +\frac{a_0 Kr}{4} \sin\phi + o(r)\,.
\qqq

We can use this result to construct $\Psi$ in the complex representation near the core. This reads 
\begin{equation}
	\label{2FCshp}
	\Psi=\frac{a_0z}{2}e^{\frac{a_0K(z-z^*)}{4}}\,.
	\end{equation}
We now use Eq. \eqref{2FCshp} to calculate the dynamics of a topological defect in a two-Frank-constant passive nematic using the Halperin-Mazenko formalism \cite{Mazenko2}.
Using this on the R.H.S of \eqref{PassEqcomplex}, we can obtain $\partial_t\Psi$ at $z=z^*=0$. This yields
\begin{equation}
	\left.\frac{a_0}{32}\left\{2a_0Ke^{a_0K(z-z^*)}{2}[8+a_0Kz(8+a_0Kz)]+a_0^3K^3zz^*-4e^{\frac{a_0K(z-z^*)}{4}}[4a_0K-4z+a_0^2z(K^2+2zz^*)]\right\}\right\vert_{z=z^*=0}=0\,.
\end{equation}
Similarly,  $\partial_t\Psi^*$ at $z=z^*=0$ also evaluates to $0$. From \eqref{velMaz}, this implies that the defect velocity $v^c_d$, calculated using the Halperin-Mazenko formalism, is $0$ for $+1/2$ defects in passive two Frank constant nematic systems if and only if we use the correct defect core shape. The correct defect shape was earlier used to calculate the dynamics of defect lines in equilibrium three-dimensional nematics with different splay and twist (but not bend) elastic constants \cite{zushi2024approach} within the Halperin-Mazenko formalism \cite{Mazenko1, Mazenko2,mazenko1999velocity} generalised for three-dimensional nematics \cite{schimming2023kinematics}.

Note that our finding here that the core angle field of a $+1/2$ passive defect with two Frank constants reduces in the $r\to 0$ limit to that of a single Frank constant defect is consistent with \cite{Schimming_Vinals}. However, we go beyond \cite{Schimming_Vinals} and analytically calculate the first order in $r$ correction to the shape, which is crucial to obtain $v^c_d$ because this depends on derivatives of $\Psi$, see eq.  \eqref{velMaz}. In contrast to \cite{Zhou_14974}, we find that the order parameter magnitude remains isotropic to leading order in $r$. However, our work cannot be directly compared to \cite{Zhou_14974} since they consider a three-dimensional system and report significant biaxiality at the core. Further, the free energy they consider is different in detail---since it contains not one but two elastic terms cubic in ${\bsf Q}$ and not one---from the two Frank constant free energy that we consider; this can also change the structure of the defect core.

\section{Relation between $+1/2$ active defect core shape and speed}
We now examine shapes of {active} $+1/2$ defects. As is well known, these have a macroscopic polar asymmetry and, therefore, self-propel due to activity. Therefore, the strategy we used in the previous sections cannot be relied upon. We assume that the defect is steadily moving at velocity ${\bf v}_d\equiv (v_d,0)$ along $\hat{{\bf x}}$ without its shape changing with time. We move to the reference frame of this defect. The governing equations now are \eqref{eq:polActTheta} and \eqref{eq:polActS} with the $\partial_t\theta$ and $\partial_tS$ on the L.H.S. replaced by $-{v_{d_\phi}\partial_\phi \theta}/{r}-v_{d_r}\partial_r \theta$ and $-{v_{d_\phi}\partial_\phi S}/{r}-v_{d_r}\partial_r S$, respectively. We note that $v_{d_r}=v_d\cos\phi$ and $v_{d_\phi}=-v_d\sin\phi$.

In this moving frame, we again use the ansatz in Eq. \eqref{eq:equilibrium-Kneq0-Ansatz+half-defect-core} for the shape of the defect at its core with $m_s>0$ and $m_\theta>0$ as $r\to 0$. This is motivated by the fact that, by definition, even in the moving frame, the order parameter amplitude has to vanish at the core and $m_\theta$ still needs to be positive in order to have a well-defined defect at $r=0$. Just like earlier, $a(\phi)$ and $b(\phi)$ are periodic functions.

Again expanding Eq. \eqref{eq:polActS} first (with the L.H.S. replaced by $-{v_{d_\phi}\partial_\phi S}/{r}-v_{d_r}\partial_r S$), at $r\to 0$, and setting the coefficients of the fastest diverging terms to $0$, we find the leading order contribution to scale as $r^{-2+m_s}$. From this, we get, as in a passive system, $m_s=1$ and $a(\phi)=a_0$ is a constant.

At next-to-leading order, we find terms that are constant in $r$. These have to be balanced by the largest remaining ones, which are of order $r^{-1+m_\theta}$. This gives
\qq
a_0^2 L_2 \cos (\phi )-4 a_0 r^{m_\theta-1} b'(\phi )+a_0 v_{d_r} =0\,.
\qqq
Hence, $m_\theta=1$ and 
\qq
b'(\phi )=
\frac{a_0}{4} L_2 \cos \phi + \frac{v_{d_r}}{4}=\frac{a_0L_2+v_d}{4}  \cos \phi \,.
\qqq
Integrating this equation gives
\qq
b(\phi )=
b_{1}+\frac{a_0 L_2+v_d}{4}\sin\phi\,.
\qqq

We then insert this form of $b(\phi )$ in the dynamics for $\theta$ (i.e., Eq. \eqref{eq:polActTheta} with the L.H.S. replaced by $-{v_{d_\phi}\partial_\phi \theta}/{r}-v_{d_r}\partial_r \theta$) and expand that for small $r$. The most diverging term is of order $1/r$, and (again using the fact that $v_{d_\phi}=-v_d\sin\phi$) this is $3b_1/r$. Since this has to vanish, $b_1=0$ and we get
\qq
b(\phi )=
\frac{a_0 L_2+v_d}{4}\sin\phi\,.
\qqq


\section{Numerical scheme for integrating dry active nematics}\label{app:nematics}
The numerical test of our analytical prediction requires the study of topological defects that are isolated. In order to do so, we performed simulations on a circular domain by imposing a topological charge at the boundary. This was done using a boundary condition that corresponds to the far-field shape of the $\pm 1/2$ topological defect that is found in passive systems with one Frank constant, which induces a topological defect of charge $\pm 1/2$ in the bulk at all times. 

To perform simulations of eq. (1) of the main text in this bounded domain, we implemented a finite element method using the MATLAB~\cite{MATLAB} partial differential equation toolbox (PDEtoolbox). The geometry is created using the \emph{PDE Modeler App (PDEMA)}. We set the characteristic length of the system as $\lambda=\sqrt{\kappa/a}=1$ and express the radius $R$ of the disk in these units. Time is expressed in units of the natural time-scale $\gamma=1/a=1$. 

	A crucial step in each finite element algorithm is the definition of a mesh that covers the domain. We created it using the  \emph{PDEtoolbox} MATLAB function which, in two-dimensional domains, uses triangular building blocks.
Unless specified otherwise, we set the maximum length of the triangle sides to be $0.5$. 
	In addition, we used a quadratic 
	 geometric interpolation via the function 
	 \textit{generateMesh()}, meaning that an extra node is 
	 added to the middle point of each side of each element.
	 To measure the core field shape of the $+1/2$ active defect 
	we needed a finer mesh grid in the core of the defect.
	 To do so we used the package \textit{refinePDEMmesh} that allows to refine the mesh in a 
	 specific region but only works with linear order elements.
	 At each iteration we calculate the average size of
	 the elements inside the refined region and repeat the process until sides have the desidered average length, which we set to $2 \times 10^{-3}$. 

	Since ${\bsf Q}$ is a symmetric and traceless rank two tensor, it only has two independent components that we take to be $Q_{xx}$ and $Q_{xy}$.
	 We consider their time evolution according to
	equation (1) of the main text with the boundary conditions
	\begin{equation}\label{eq:BC_Q}
		\begin{cases}
			Q_{xx}(R,\phi) = \dfrac{1}{\sqrt{2}}\cos\left(2\theta(R,\phi)\right)\,,\\
			Q_{xy}(R,\phi) = \dfrac{1}{\sqrt{2}} \sin\left(2\theta(R,\phi)\right)\,,
		\end{cases}
	\end{equation}
	where, for the $\pm 1/2$ defect:
	\begin{equation}
		\theta(R,\phi) = \pm\dfrac{\phi}{2} + \frac{\pi}{2}\,.
	\end{equation}
As initial conditions, we used ones that possess a defect located in the origin and are compatible with \eqref{eq:BC_Q}. This is achieved by imposing at $t=0$:
	\begin{equation}\label{eq:IC_Q}
		\begin{cases}
			Q_{xx}(r,\phi) = \dfrac{r}{\sqrt{2}R} \cos\left(\pm\phi+\pi\right)\,\\
			Q_{xy}(r,\phi) = \dfrac{r}{\sqrt{2}R} \sin\left(\pm\phi+\pi\right)\,.
		\end{cases}
	\end{equation}
	
	 \begin{figure}
	\begin{centering}
		\includegraphics[width=0.5\columnwidth]{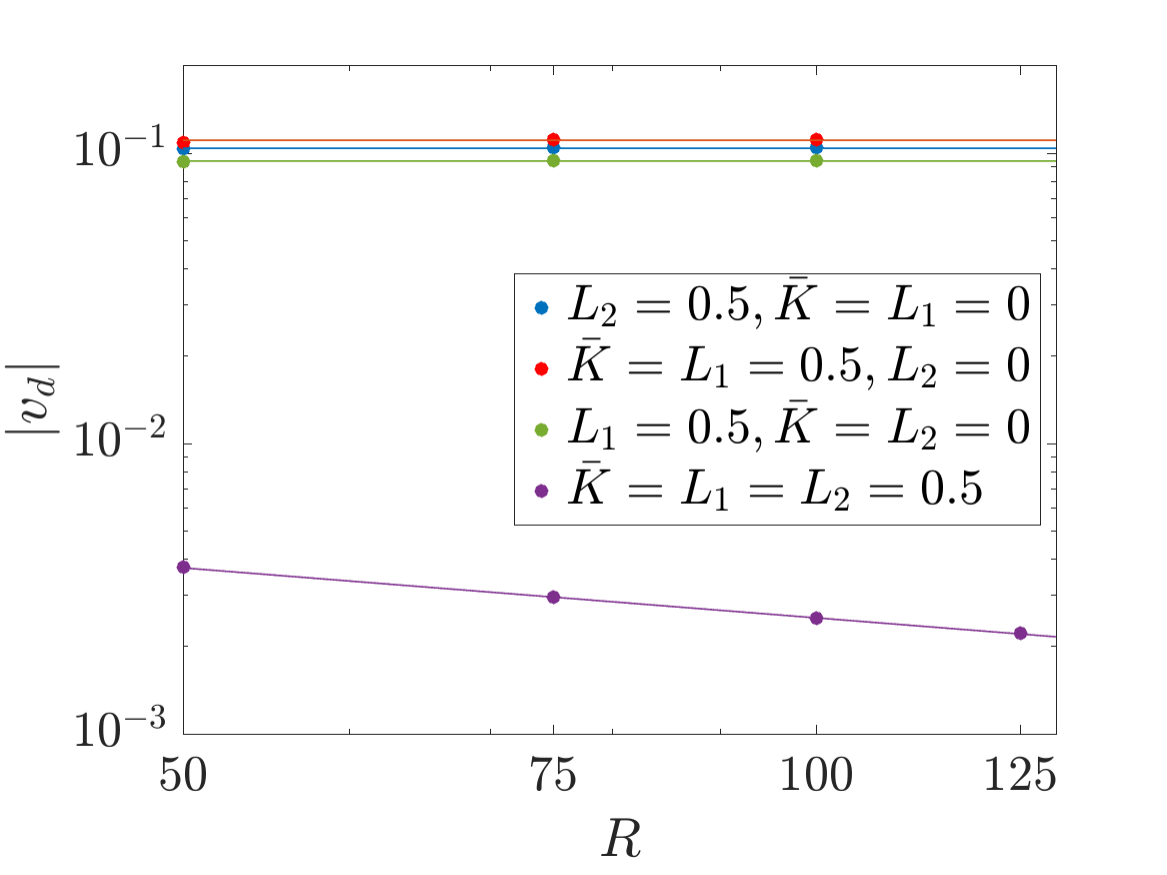}
		\par\end{centering}
	\caption{Speed of the $+1/2$ defect as measured in simulations in three active cases and the passive liquid crystal with two Frank constants as indicated in the legend as found for different system sizes (lines are guides to the eye). In all active cases, irrespectively of whether $L_2\neq 0$, $v_d$ is independent of system-size, while it decreases to zero with increasing $R$ in the passive case as expected.}\label{Fig-SM-1}
\end{figure}

	The \emph{PDEToolbox} discretizes the problem in space via the Galerkin method and provides a non-linear differential equation solver with adaptive time stepping scheme that we used with default parameters reported in the documentation~\cite{MATLAB-documentation}. 
	The solver only accepts equations for the vector~$\mathbf{u} = (Q_{xx},Q_{xy})^T$ expressed in the divergence form:
	\begin{equation}\label{eq:divergenceFormToolboxEq}
 \mathbf{d}\partial_{t}\mathbf{u} - \nabla \cdot (\mathbf{c} \otimes \nabla)\mathbf{u} + \mathbf{au} = \mathbf{f}
	\end{equation}
	The coefficients $ \mathbf{a}, \mathbf{c},  \mathbf{d},  \mathbf{f}$ are matrices which can depend on $\mathbf{u}$ or its first spatial derivative. 	
	A tedious but simple calculation shows that eq. (2) can be rewritten in the form of \eqref{eq:divergenceFormToolboxEq}. The result for the case $K_1=K_2=\bar{K}$, which is the case we consider all through the Letter, is
	\begin{equation}\label{eq:dCoeff}
		\mathbf{d} = \begin{pmatrix}
			-1 & 0 \\
			0 & -1 \\
		\end{pmatrix}\,,
	\end{equation}
	\begin{small}
		\begin{equation}\label{eq:aCoeff}
			\mathbf{a} = \begin{pmatrix}
				a - 2 b\left( Q_{xx}^{2} + Q_{xy}^{2} \right) & 0 \\
				0 & a-2b\left( Q_{xx}^{2} + Q_{xy}^{2} \right) 
			\end{pmatrix}\,,
		\end{equation}
	\end{small}
	\begin{equation}\label{eq:cCoeff}
		\begin{small}
			\mathbf{c} = \begin{pmatrix}
				-\kappa-L_{1}Q_{xx} & -L_{1}Q_{xy} & 0 & 0 \\
				-L_{1}Q_{xy} & -\kappa+L_{1}Q_{xx} & 0 & 0 \\
				0 & 0 & -\kappa-L_{1}Q_{xx} & -L_{1}Q_{xy}\\
				0 & 0 & -L_{1}Q_{xy} & -\kappa+L_{1}Q_{xx} 
			\end{pmatrix}\,,
		\end{small}
	\end{equation}
	\begin{equation}\label{eq:fCoeff}
		\begin{split}
			\text{f}_{xx} =&
			\left(\bar{K}/2+L_{1}-L_{2}\right) \left(\partial_{x}Q_{xx}\right)^{2} 
			- \left(\bar{K}/2+L_{1}-L_{2}\right) \left(\partial_{y}Q_{xx}\right)^{2}+ (\bar{K}/2)\left(\partial_{x}Q_{xy}\right)^{2}\\
			&- (\bar{K}/2)\left(\partial_{y}Q_{xy}\right)^{2} + \left(L_{1}-L_{2}\right)\left(\partial_{x}Q_{xy}\right)\left(\partial_{y}Q_{xx}\right) 
			+\left(L_{1}-L_{2}\right)\left(\partial_{y}Q_{xy}\right)\left(\partial_{x}Q_{xx}\right)\,,\\
			\text{f}_{xy} = & 
			\bar{K}\left(\partial_{x}Q_{xx}\right)\left(\partial_{y}Q_{xx}\right) 
			+\left(\bar{K}+2L_{1}+2L_{2}\right)\left(\partial_{x}Q_{xy}\right)\left(\partial_{y}Q_{xy}\right)\\
			&+ \left(L_{1}-L_{2}\right)\left(\partial_{x}Q_{xx}\right)\left(\partial_{x}Q_{xy}\right) 
			- \left(L_{1}-L_{2}\right)\left(\partial_{y}Q_{xx}\right)\left(\partial_{y}Q_{xy}\right)\,.
		\end{split}
	\end{equation}
	
	Before addressing the defect shape of the active system in eq. (1) of the main text (the results from which are described in the main text), we tested the algorithm on the one Frank constant model, obtained from equation (1) of the main text by setting $a=b=\kappa=1$ and $K_{1}= K_{2}=L_{1}=L_{2}=0$.
	 For this case the defect shape is analytically known both for $\pm 1/2$ defects to be 
\begin{align}
		&S(r) = \sqrt{2}-\dfrac{1}{\sqrt{2}r^{2}} +o(1/r^2),&r\rightarrow\infty \label{eq:farField1FC}\\
		&S(r) = a_0 r+o(r),&r\rightarrow 0 \label{eq:coreField1FC}
	\end{align}
	where $a_0\simeq 0.82$ up to an error of $3\%$ as it follows from a Pad\'e expansion~\cite{pismen1999vortices}. 
	Our simulations (performed for $R=50$ and integrating over time up to $t_{\mathrm{f}}=100$, data not shown)
	 perfectly reproduced both the near and far field results in eq. \eqref{eq:farField1FC} and \eqref{eq:coreField1FC}, obtained by respectively measuring $S$ at $r<1$ and and $10<r<R$. We further checked measuring $\theta$ as a function of $r$ that $b(\phi)$ exactly vanishes in this case as expected. 

\subsection{Drifting of the $+1/2$ defect in passive systems}
In the main text we have discussed that the $+1/2$ defect self-propels in all active systems. In our simulations, the $+1/2$ defect drifts even in the passive system with two Frank constants, although with a very small speed. Furthermore Fig. \ref{Fig-SM-1} shows that the resulting drift velocity decreases when increasing the system size. This is a fundamental difference with respect to the case of active systems, where we have shown (see the main text) that the self-propulsion speed is well defined when increasing system-size. 
The reason for which the $+1/2$ defect drifts in our passive simulations is a boundary effect analogous to the Peach-Koehler force~\cite{long2024applications}, due to imposing at the boundary the nematic order corresponding to the single Frank constant defect, which is not the far field solution of the two Frank constant defect. The natural expectation is that the defect speed decreases as $1/R$ for large system-sizes~\cite{eshelby2006force,cermelli2002evolution}; however, likely due to computational limitations in system-size, we could not observe this scaling cleanly and we instead observed the defect speed decreasing approximately as $1/R^{0.6}$ (but with indication of fast decrease at the larger $R$ simulated). The $-1/2$ defect does not drift because the angular deformation at the boundary due to the imposition of the nematic order has three-fold symmetry.

\subsection{Checking for the effects of the boundary}
\begin{figure}[h]\center
	\centering
	\includegraphics[width=0.5\linewidth]{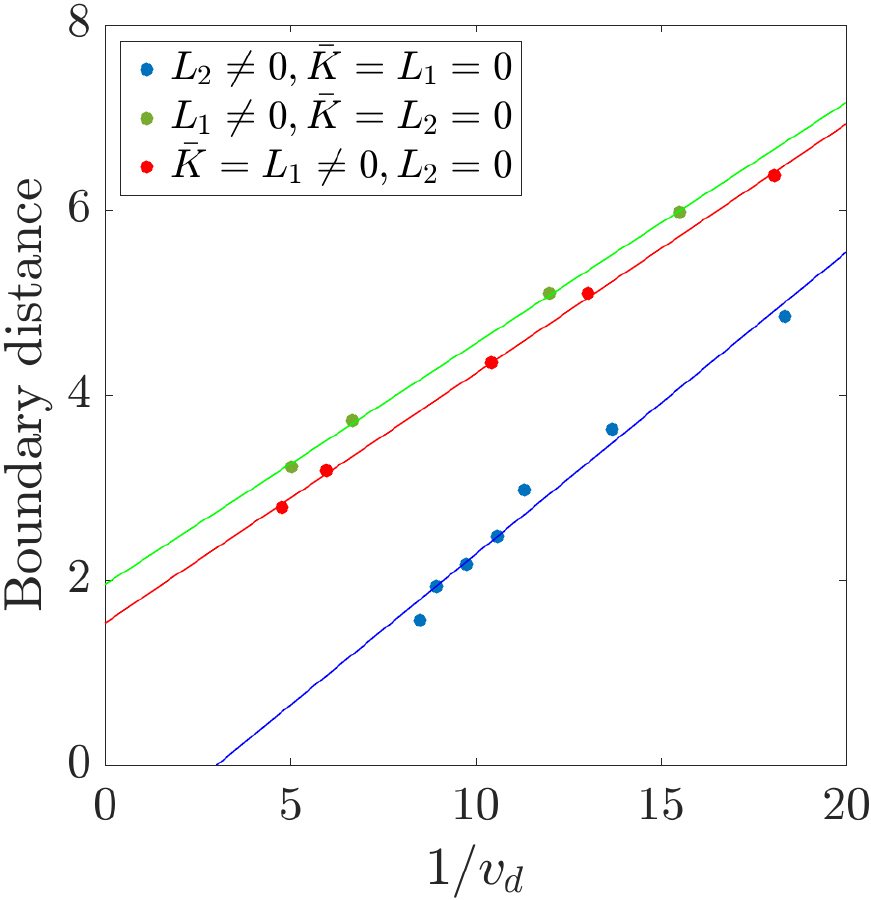}
	\caption{Distance from boundary at which $+1/2$ active, self-propelling defects stop as a function of $1/v_d$ upon varying $L_i,K_i$, $i=1,2$. }
	\label{defemov}
	\end{figure}

We now perform a second numerical check to ascertain the effect of boundaries. We expect on dimensional grounds that the distance at which active defects stop away from the boundary scales as $1/v_d$ (going back to dimensional variables, $\kappa/|{\bf v}_d|$  is a lengthscale, where $\kappa$ is the coefficient of the $\nabla^2{\bsf Q}$ in the equation of motion). We find that the motile defects in our simulations indeed obey this scaling as displayed in Fig. \ref{defemov}. 

\section{Model of active nematic with coupled equations for ${\bsf Q}$ tensor and the velocity field}
``Wet'' active nematic often refers to models describing an incompressible fluid in which the total momentum is conserved~\cite{RMP}. It was shown that an ordered nematic phase does not exist \cite{Aditi}, because it is generically destroyed by an instability caused by long-range interactions induced by momentum conservation and incompressibility \cite{Aditi, Ano_2d}. Therefore, one can talk of defects in that case only at scales smaller than the minimal length-scale at with the instability sets in; this is proportional to the square root of the Frank elasticity divided by activity. In contrast, when active fluids are in contact with substrates, such that the total momentum is not conserved, the nematic state is not necessarily unstable \cite{Raphael, AM_PNAS}. Further, active nematic order can also exist at fluid interfaces \cite{Ano_2d}. However, these phases are qualitatively distinct from the dry system we consider in this article: incompressibility in the first case and interactions mediated by the bulk fluid in the second induce long-range interactions.  We now show that starting from a model of ${\bf Q}$ tensor and a frictionally-screened  velocity field ${\bf v}$, we obtain a nematic model of the sort that we consider (we do not consider density fluctuations). This model is expected to be valid when the constrains of incompressibility and momentum conservation are relaxed which happens, for example, when the active nematics is on a porous substrate that is permeable to fluid.

We consider a model similar to the one introduced in~\cite{Suraj_def}, although the non-linearities of our dry model in eq. (1) of the main text would be obtained also starting from a significant simpler model, but without their coefficients being independent. 
The dynamics of the ${\bsf Q}$ tensor coupled to the velocity field is given by
\begin{align}
	\label{Qeqnfl}
	\partial_tQ_{ij}&+v_k\partial_kQ_{ij}+\frac{1}{2}[Q_{ik}(\partial_kv_j-\partial_jv_k)+Q_{kj}(\partial_kv_i-\partial_iv_k)]=\nonumber\\
	&=-\frac{\lambda}{2}(\partial_iv_j+\partial_jv_i-\delta_{ij}\partial_kv_k)+\lambda_2Q_{ij}\partial_kv_k+\lambda_v\left(v_iv_j-\frac{1}{2}\delta_{ij}v_kv_k\right)+ H_{ij}\,,\\
		\Gamma v_i&=\zeta\partial_jQ_{ij}+\zeta_2Q_{ik}\partial_lQ_{kl}+\zeta_3\partial_i(Q_{kl}Q_{kl})-H_{lk}\partial_iQ_{lk}\nonumber\\
		&-\partial_j(Q_{il}H_{lj}-H_{il}Q_{lj}+\lambda H_{ij})-\lambda_2\partial_i(H_{lj}Q_{lj})-\lambda_vv_jH_{ij}	\label{genveqn}
\end{align}
where $H_{ij}=-[\delta F/\delta Q_{ij}]^{ST}$ (we have also set a kinetic coefficient to $1$ by a suitable redefinition of $t$), and the superscript $ST$ denotes the trace-removed and symmetrised part of a tensor. 
When $\zeta=\zeta_2=\zeta_3=0$, this pair of equations describe an equilibrium dynamics (once the noise with proper correlation has been included). In this case, replacing $v_i$ in \eqref{Qeqnfl} using \eqref{genveqn}, we obtain an equation of the form $\partial_tQ_{ij}=\Gamma_{Q_{ijkl}}H_{kl}$, where $\Gamma_{Q_{ijkl}}$ depends on both ${\bsf Q}$ and $\nabla$. Notice also that other terms in eq. \eqref{Qeqnfl} and \eqref{genveqn}, to the same order in $\bf v$ and $\bf Q$, are possible: we do not consider all of them here for simplicity, as they are irrelevant to our argument.

We now eliminate the velocity and obtain an equation for $Q_{ij}$ where we retain terms with only two derivatives and ${\bsf Q}$s:
\begin{multline}
	\label{sing_eq}
	\partial_tQ_{ij}+\frac{\zeta}{\Gamma}\partial_lQ_{kl}\partial_kQ_{ij}+\frac{\zeta}{2\Gamma}[Q_{ik}(\partial_k\partial_lQ_{jl}-\partial_j\partial_lQ_{kl})+Q_{jk}(\partial_k\partial_lQ_{il}-\partial_i\partial_lQ_{kl})]=-\frac{\lambda\zeta}{2\Gamma}(\partial_i\partial_lQ_{jl}+\partial_j\partial_lQ_{il}-\delta_{ij}\partial_k\partial_lQ_{kl})\\-\frac{\lambda\zeta_2}{2\Gamma}[\partial_i(Q_{jk}\partial_lQ_{kl})+\partial_j(Q_{ik}\partial_lQ_{kl})-\delta_{ij}\partial_k(Q_{km}\partial_lQ_{lm})]-\frac{\lambda\zeta_3}{2\Gamma}[\partial_i\partial_j(Q_{kl}Q_{kl})+\partial_j\partial_i(Q_{kl}Q_{kl})-\delta_{ij}\partial_k\partial_k(Q_{lm}Q_{lm})]\\+\frac{\lambda_2\zeta}{2\Gamma}Q_{ij}\partial_k\partial_lQ_{kl}+\frac{\lambda_v\zeta^2}{\Gamma^2}\left(\partial_lQ_{il}\partial_mQ_{jm}-\frac{1}{2}\delta_{ij}\partial_lQ_{kl}\partial_mQ_{km}\right)+\Gamma_{Q_{ijkl}}H_{kl}\,.
\end{multline}
where we have kept track explicitly only of the terms involving $\zeta_{1,2,3}$ in \eqref{genveqn}, i.e. we left implicit the term $\Gamma_{Q_{ijkl}}H_{kl}$. 
Eq. \eqref{sing_eq} is now a closed equation for $Q_{ij}$ and explicitly contains the non-linearities proportional to $K_1$ and $L_2$ in eq. (1) of the main text. The technique developed in this Letter could be extended to obtain the shape and motion of defects implied by Eq. \eqref{sing_eq}.

As we have shown, the shape of defects is crucially linked to their properties such as the self-propulsion. This was best shown in the model in eq. (1) of the main text, that has a transparent limit in which it reduces to a passive system with two Frank constants; the technique developed in this Letter could be extended to obtain the shape and motion of defects implied by Eq. \eqref{sing_eq}, but we do not explore this avenue further here. However, there is something to learn from eq.\eqref{sing_eq}: It is often suggested that the direction of defect self-propulsion is controlled by the contractile or extensile character of the active force density \cite{Doostmohammadi_rev, Suraj_rev,Sagues_bk}, 
which is determined solely by the sign of $\zeta$. The plethora of free parameters in \eqref{sing_eq}, all of which enter the expression for defect motion shows that this is unlikely to be the case, at least in the dry limit. This was already observed in~\cite{Suraj_def}.

\section{Brief comments on defects in wet active systems}
Incompressible or momentum-conserving flow introduces spatially non-local terms in the dynamics of the ${\bsf Q}$ tensor even in the non-inertial limit relevant for most biological materials. We will not discuss the defect shapes that arise in the presence of such non-local interactions; however, in this subsection we will argue that defect shape modifications are equally important in the dry and wet cases.

For this, we first show that even in passive wet nematic fluid the defect speed is predicted not to vanish, exactly as discussed in the main text for the dry case, if shape modifications are not taken into account. The model we consider is a passive, incompressible nematic fluid, with different elastic cost for splay and bend perturbations, that conserves momentum. This corresponds to setting vanishing activity ($\zeta=\zeta_2=\zeta_3=0$) in eq. \eqref{genveqn}, and to setting $\lambda_2=\lambda_v=0$ in eq. \eqref{Qeqnfl}. The last two choices respectively enforce incompressibility and Galileian invariance.  In complex notation introduced in Sec. \ref{CompEq}, we have
\begin{equation}
	\label{Psieqvel}
	\partial_t\Psi=-[v\partial +v^*\partial^*]\Psi+{\Psi}[\partial v-\partial^*v^*]-{\lambda}\partial^*v+H_\Psi\,,
\end{equation}
\begin{equation}
	\label{compvel}
	-2\eta\partial\partial^*v=\lambda\partial H_\Psi-\partial^*(\Psi^*H_\Psi-\Psi H_\Psi^*)-(H_\Psi\partial^*\Psi^*+H^*_\Psi\partial^*\Psi)-\partial^*\Pi\,,
\end{equation}
where
\begin{equation}
	H_\Psi=(1-2|\Psi|^2)\Psi+4\partial\partial^*\Psi+K[2\Psi^*{\partial^*}^2\Psi+\partial^2(\Psi^2)]\,,
\end{equation}
and  $\Pi$ is the pressure that enforces the incompressibility constraint $\partial v+\partial^*v^*=0$. 

We now use the single Frank constant defect shape and obtain the velocity of a $+1/2$ defect implied by \eqref{Psieqvel} and \eqref{compvel} using the Halperin-Mazenko method. For a defect at the origin, this yields
\begin{equation}
	\label{defspwrng}
	v^c_d=-\frac{2}{a_0}\partial_t\Psi(z,z*=0)=-Ka_0+[v+\lambda\partial^*v]_{z,z*=0}
\end{equation}
Thus, the flow in this model leads to additional terms involving the fluid velocity with respect to the dry case. These terms can be calculated within the approximation used in \cite{Angheluta2}. However, such calculation is not needed to conclude that eq. \eqref{defspwrng} is wrong: As the system is at equilibrium, we must have $v^c_d=0$. Yet, this is impossible for arbitrary parameters because the terms in the square bracket must be proportional to $1/\eta$ while the first term in eq. \eqref{defspwrng} is not. This implies that to obtain even qualitatively accurate results for the defect speed, defect shape change must be taken into account in wet systems as well.

We now provide an example in an active system in which defect shape is expected to make a qualitative difference. What follows corrects an incorrect statement---that a force density of a specific type \emph{cannot} lead to defect self-propulsion of $+1/2$ defects---made in \cite{AM_PNAS} and further shows the importance of accounting for defect shape even in wet active systems.
 Consider an incompressible active fluid on a substrate in which the coefficient of the standard active force density $\nabla\cdot{\bsf Q}$ vanishes, but in which there is a different active force density $\zeta_2Q_{ik}\partial_lQ_{lk}$ \cite{AM_PNAS}. The question is again whether $+1/2$ defects self-propel. If we plug in the form that ${\bsf Q}$ assumes in the far-field for a $+1/2$ defect in an equilibrium nematic with a single Frank constant (as is done for calculating the defect motion in \cite{Angheluta2, Giomi_def}), we find that 
\begin{equation}
\zeta_2{\bsf Q}\cdot(\nabla\cdot{\bsf Q})=\zeta_2\frac{\hat{{\bf r}}}{2r}\,,
\end{equation}
in polar coordinates. The two-dimensional curl of this vanishes as noted in \cite{AM_PNAS} and, therefore, it can be expressed as a pure gradient. This implies that if the true shape of the $+1/2$ active defect were indeed the same as the one in passive single Frank constant systems, this force density could not lead to any flow (due to incompressibility), and thus to no self-propulsion of $+1/2$ defects. However, this conclusion does not hold if the defect shape deviates from the single Frank constant one. To see this, we insert the perturbatively calculated far-field shape of a $+1/2$ defect in a two Frank constant nematic in the expression for force density. Using only the first order in $K$ correction obtained in \eqref{+halfpassiveshape}, we obtain 
\begin{equation}
	\label{Eq2actfrc}
	\zeta_2{\bsf Q}\cdot(\nabla\cdot{\bsf Q})=\zeta_2\left[(4+3\sqrt{2}K\cos\phi)\frac{\hat{{\bf r}}}{8r}\right]\,,
\end{equation}
whose two-dimensional curl is $3\zeta_2K\sin\phi/(4\sqrt{2}r^2)$. This force density cannot be compensated by a pressure in an incompressible system and leads to flows. Further, given that the curl of Eq. \eqref{Eq2actfrc} has a similar form to the curl of the usual active force density due to a $+1/2$ single Frank constant defect shape  (for comparison, the curl of the usual active force density $\zeta\nabla\cdot{\bsf Q}$ due to a $+1/2$ single Frank constant defect shape is given by $\zeta\sin\phi/(\sqrt{2}r^2)$), we expect the $+1/2$ defect to self-propel. Of course, this qualitative result should hold even when the defect shape is not given by a two-Frank-constant passive nematic theory; in other words, even active shape modifications away from a single Frank constant defect shape should generically lead to the self-propulsion of $+1/2$ defects in incompressible active nematics even when the only active force density is $\propto{\bsf Q}\cdot(\nabla\cdot{\bsf Q})$.

\end{document}